\documentclass[epj]{svjour}
\usepackage{verbatim} 
\usepackage{graphicx}
\usepackage{dcolumn}

\usepackage{amstext,amsmath,amssymb,amsfonts}
\usepackage{bm}

\usepackage{txfonts}
\usepackage{mathtools}

\usepackage{hyperref}

\begin{document}

\title{Measures of azimuthal anisotropy in high-energy collisions} 
\dedication{In memory of Art Poskanzer}

\author{Jean-Yves Ollitrault}
\mail{jean-yves.ollitrault@ipht.fr}
\institute{Universit\'e Paris Saclay, CNRS, CEA, Institut de physique th\'eorique, 91191 Gif-sur-Yvette, France}

\date{\today}

\abstract{
Azimuthal anisotropy is a key observation made in ultrarelativistic heavy-ion collisions. 
This phenomenon has played a crucial role in the development of the field over the last two decades. 
In addition to its interest for studying the quark-gluon plasma, which was the original motivation, it is sensitive to the properties of incoming nuclei, in particular to the nuclear deformation and to the nuclear skin. 
The azimuthal anisotropy is therefore of crucial importance when relating low-energy nuclear structure to high-energy nuclear collisions. 
This article is an elementary introduction to the various observables used in order to characterize azimuthal anisotropy, which go under the names of $v_2\{2\}$, $v_3\{2\}$, $v_2\{4\}$, etc. 
The intended audience is primarily physicists working in the field of nuclear structure. 
}

\maketitle

\section{Why heavy-ion colliders?}
\label{sec:intro}

\begin{figure}[tb]
\begin{center}
\includegraphics[width=0.61\linewidth]{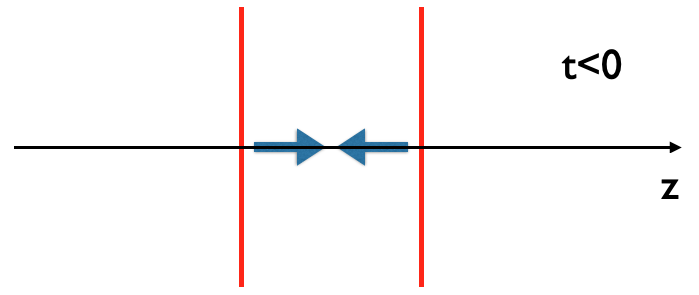}
\includegraphics[width=0.62\linewidth]{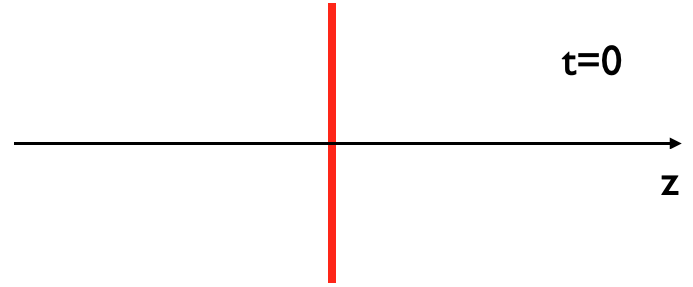} 
\includegraphics[width=0.61\linewidth]{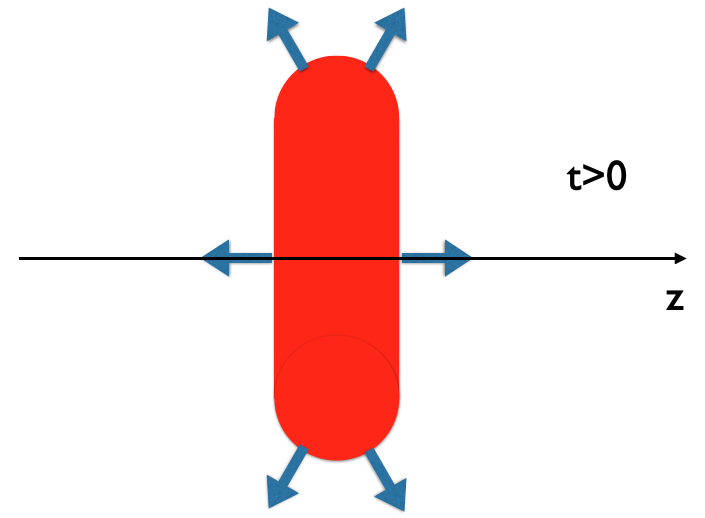} 
\end{center}
\caption{Schematic side view of a central, symmetric nucleus-nucleus at the RHIC or LHC collider in the laboratory frame.  
When $t<0$ (top), the nuclei move towards each other along the collision axis $z$. 
The relativistic length contraction along the direction of motion is so strong (by a factor up to $\sim 100$ at RHIC, $\sim 3000$ at the LHC) that spherical nuclei appear as flat disks. 
The collision itself occurs over a very short time scale, which is conventionally chosen as the origin $t=0$ (middle). 
Matter created in the collision then expands in all directions at a velocity close to that of light. 
Therefore, the region it fills at time $t>0$ is depicted as a rectangle of width $\simeq 2 c t$, capped with two semicircles of radius $\simeq ct$ (bottom). 
}
\label{fig:schema}
\end{figure}

It is counter-intuitive that one might learn something about the internal structure of atomic nuclei by smashing them against one another at energies so high, that very little seems to remain of them after the collision, beyond the total charge and baryon number, which are conserved throughout the collision process. 
It actually turns out that the higher the collision energy, the more precise the information one can get, for reasons I now outline.  

The first important point is that the time scale over which the two nuclei interact is extremely short. 
At the RHIC and LHC colliders, two beams of atomic nuclei (fully-ionized atoms, from which all electrons have been stripped) are accelerated in opposite directions at identical speeds, very close to the speed of light $c$. 
Each nucleus is flattened along its direction of motion. 
Special relativity teaches us that the Lorentz contraction factor is $\gamma\simeq\sqrt{s_{NN}}/2 m_Nc^2$, where $\sqrt{s_{NN}}$ denotes the energy per nucleon pair in the center-of-mass frame, which is the standard measure of the collision energy, and $m_N=0.9315$~GeV$/c^2$ is the atomic mass unit.  
The contraction factor $\gamma$ goes up to $107$ at the RHIC collider, and is currently around $2900$ at the LHC~\cite{Busza:2018rrf}, so that nuclei appear as extremely flat disks, as illustrated in Fig.~\ref{fig:schema}.\footnote{This comment applies to the spatial image of the nucleus. As we shall discuss in the next paragraph, however, nucleons are made of partons, and the longitudinal extension of their wavefunction can be significantly larger.} 
The time it takes for them to pass through each other is $t_{\rm coll}=2R/\gamma c$, where $R$ is the nuclear radius. 
For Pb+Pb collisions at the LHC, where $R\simeq 7$~fm, one finds $t_{\rm coll}\simeq 2\times 10^{-26}$~s, shorter by orders of magnitude than {\it all\/} the times scales of nuclear physics.\footnote{The motivation for building ion colliders (RHIC, then LHC), with two beams accelerated in opposite directions, is that $\sqrt{s_{NN}}$ is much larger than in a fixed-target experiment, by a factor $\sqrt{2\gamma}$ in the limit $\gamma\gg 1$. In fixed-target experiments at the LHC,  $\sqrt{s_{NN}}$ is reduced by a factor $\sim 75$ relative to the collider mode~\cite{Hadjidakis:2018ifr}.}

Consequences are twofold. 
First, the Lorentz contraction effectively projects nuclei onto the transverse plane. 
The intuition from classical physics is that collisions occur {\it sequentially\/} in the longitudinal direction, with the nucleons on the front line interacting first, and nucleons on the last line interacting last. 
But quantum mechanics prohibits such an ordering. 
The wavefunctions of nucleons consist of a large number of partons, mostly gluons, each of which carries a small fraction $x_{\rm Bj}$ (the so-called ``Bjorken $x$''~\cite{Bjorken:1968dy}) of the momentum $P\simeq \sqrt{s_{NN}}/2c$ of the nucleon. 
According to Heisenberg's principle, the longitudinal extension of the wavefunction of a parton is $\delta z_{\rm parton}\sim \hbar/(x_{\rm Bj}P)$. 
This must be compared with the contracted length of the nucleus, $\delta z_{\rm nucleus}\sim 2R/\gamma$. 
A back-of-the-envelope calculation shows that if $x_{\rm Bj}<10^{-2}$, which is where most of the partons sit at very high energies, $\delta z_{\rm parton}>\delta z_{\rm nucleus}$. 
This implies that in the laboratory frame, the wavefunctions of nucleons fully overlap in the longitudinal direction. 
Therefore, all the nucleon-nucleon collisions happen {\it simultaneously\/}. 
The only information that remains from the nuclear structure is integrated over the longitudinal coordinate $z$. 
If $\rho(x,y,z)$ denotes a physical quantity characterizing the local density of the nucleus at the time of the collision (from now on, $x$ denotes a spatial cartesian coordinate, not to be confused with the Bjorken $x$ above), then the collision is only sensitive to the thickness function $T_A(x,y)\equiv\int_{-\infty}^{+\infty}\rho(x,y,z)dz$. 

Second, the collision can be thought of as a snapshot, which captures the positions of nucleons within nuclei, as well as the internal structure of these nucleons, at the time when the collision occurs. 
In this respect, the collision acts as a measurement, in the quantum sense. 
For this reason, high-energy collisions differ fundamentally from low-energy experiments.  
Quantum mechanics plays an essential role in describing nuclear structure. 
Because of quantum mechanics, a phenomenon as intuitive as the ellipsoidal deformation of the $^{238}$U nucleus in its ground state is not easy to formulate. 
One considers it to be deformed in a specific frame, called the ``intrinsic frame''. 
But $^{238}$U in its ground state has no rotational motion (it is a $j=0$ state in the language of quantum mechanics), which implies that its wave function is spherically symmetric in the laboratory frame. 
This spherical wave function is written as a linear superposition of ellipsoidal states with all possible orientations. 
In a high-energy collision, the wave function collapses and one measures a single orientation. 
Intuitively, the nucleus is seen in its intrinsic frame.  

The major complication of a heavy-ion collision, respective to nuclear structure, is that it involves two nuclei, instead of one.  
What one sees at the end of the day is a number of particles emitted in the collision. 
Over the years, evidence has accumulated that the density of matter produced in the collision, which determines the emission of particles, is proportional to the product of the nuclear profiles\footnote{This prescription was initially inspired by descriptions using high-energy QCD to model the collision~\cite{Kajantie:1987pd,Krasnitz:1998ns,Eskola:1999fc,Eskola:2001bf,Schenke:2012wb}.}  of the incoming nuclei, $T_A(x,y)T_B(x,y)$~\cite{Moreland:2014oya,Nagle:2018ybc,Schenke:2020mbo}. 
This lore provides a direct link between high-energy collisions and nuclear structure~\cite{Giacalone:2023hwk}. 

The main reason why high energies are useful is that the higher the energy, the larger the number of particles created in the collision. 
In a head-on collision (zero impact parameter, as in Fig.~\ref{fig:schema}) at ultrarelativistic energies, created particles (in majority pions) largely outnumber the initial number of nucleons (which is  $2A$ for a collision between two identical nuclei of mass number $A$), by a factor $\sim 20$ at the top RHIC energy~\cite{Braun-Munzinger:2003pwq}, and by a factor $\sim 85$ at the LHC~\cite{ALICE:2016fbt}.\footnote{ALICE evaluates the average number of charged particles for collisions in the 0-5\% centrality window. One must in addition take into account that $\sim \frac{1}{3}$ of the produced particles are neutral, and that the average multiplicity in $b=0$ collisions is $\sim 11\%$ larger than in the considered centrality window, which consists roughly of collisions with $b<3.5$~fm~\cite{Das:2017ned}. }
This implies in practice that more information is available in every collision. 

\begin{figure}[tb]
\begin{center}
\includegraphics[width=.8\linewidth]{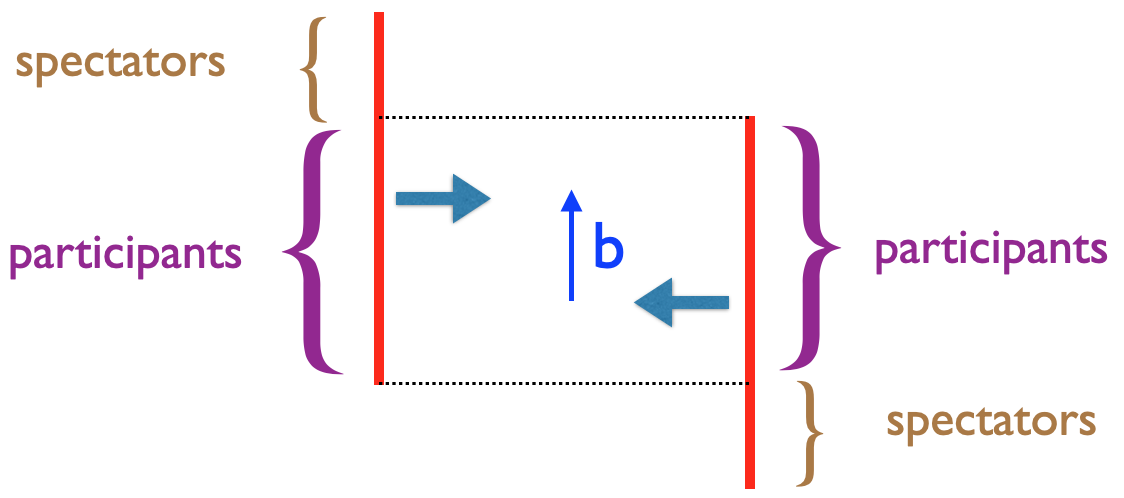}
\end{center}
\caption{Schematic side view of a symmetric nucleus-nucleus collision at impact parameter $b$. 
The nucleons of incoming nuclei are classified as participants or spectators, depending on whether or not they collide.   
}
\label{fig:bside}
\end{figure}

A concrete example of the usefulness of a larger multiplicity is the determination of the impact parameter $b$ of the collision from experimental data, which becomes significantly more precise as the collision energy increases. 
At ultrarelativistic energies, the impact parameter is a well-defined quantity, in the sense that its quantum uncertainty $\delta b=\hbar/P$, where $P$ is the momentum of the nucleus, is negligible. 
For Pb+Pb collisions at the LHC, $\delta b\simeq 4\times 10^{-7}$~fm, negligible compared to the nuclear diameter $2R\simeq 14$~fm. 
Despite being well defined, the impact parameter cannot be measured directly in experiment, and it is estimated using the multiplicity of the event, as I now explain. 
The only nucleons which interact are those which overlap with the other nucleus at the time of impact (Fig.~\ref{fig:bside}). 
They are referred to as participants or wounded nucleons~\cite{Miller:2007ri}, while the other nucleons are called spectators. 
The number of particles produced in the collision is proportional to the number of wounded nucleons (more precisely, to the number of wounded quarks within colliding nucleons~\cite{Moreland:2014oya,Eremin:2003qn,Bialas:2006kw,PHENIX:2013ehw,Loizides:2016djv}). 
Larger impact parameter implies fewer wounded nucleons and smaller multiplicity. 

This property is used in the following way. 
One records a large number of collisions (called ``events'') and one sorts them by ascending values of the multiplicity (or an observable proportional to it~\cite{ALICE:2013hur}). 
This amounts to sorting them by decreasing values of $b$, and this is referred to as a ``centrality classification'' by heavy-ion experiments. 
The centrality is usually given in percentiles, the 5\% most central collisions being defined as the 5\% with the highest multiplicities. 
This classification does not perfectly reflect the ordering in $b$, because two collisions with the same $b$ may have slightly different multiplicities. 
These multiplicity fluctuations are smaller when more particles are seen, so that the higher the energy, the better the centrality resolution~\cite{Das:2017ned}. 
The root-mean-square error on the centrality for central collisions of heavy nuclei is typically $2\%$ at the top RHIC energy and $1\%$ at the LHC,\footnote{Multiplicity fluctuations at fixed $b$ are partly due to statistical (Poisson) fluctuations, which result in a standard deviation of $\sqrt{N}$ if the average multiplicity is $N$. Relative statistical fluctuations are reduced if the detector acceptance is larger, so that it sees more particles. Statistical fluctuations dominate at lower energies, but they contribute by less than $20\%$ to the variance at the LHC~\cite{Yousefnia:2021cup}, so that upgrading detectors should not significantly improve the centrality resolution.} so that increasing the collision energy results in a significant improvement. 

The larger multiplicity is even more crucial for measuring azimuthal anisotropies, which are the main topic of this article. 
They are deformations of the azimuthal distribution of outgoing particles. 
The physics underlying this phenomenon, as well as the quantities used to quantify it, present several analogies with the physics of nuclear deformation, with the momenta of outgoing particles playing the role of nucleon positions.  
Azimuthal anisotropy is generated by collective motion of outgoing particles, in the same way as the nuclear deformation is a collective effect, hence the term ``anisotropic flow'' coined to describe this phenomenon~\cite{Poskanzer:1998yz}. 
An important difference with low-energy nuclear physics is the projection onto the transverse plane, resulting from Lorentz contraction, as a consequence of which only angle, the azimuthal angle $\phi$,  is relevant. 
While the nuclear deformation is defined by expanding the nuclear shape in spherical harmonics $Y_{l,m}(\theta,\phi)$, the azimuthal distribution of outgoing particles in heavy-ion collisions is expanded in a simple Fourier series, a superposition of $\exp(-in\phi)$, where $n$ is the harmonic order. 
In the same way as the nuclear deformation is quantified by parameters $\beta_{l,m}$, which are the coefficients in front of of the corresponding spherical harmonics, the azimuthal anisotropy is quantified by Fourier coefficients $v_n$, which will be defined below. 
Elliptic flow $v_2$ is similar to the quadrupole deformation $\beta_{2,0}$, triangular flow $v_3$ is similar to the octupole deformation $\beta_{3,0}$, etc. 


The historical development of this field also bears some analogy with that of nuclear deformation. 
I first discuss elliptic flow in non-central collisions (Sec.~\ref{sec:v2}), which is formally analogous to a static deformation. 
Then I present the modern picture of azimuthal anisotropy (Sec.~\ref{sec:ridge}), which is based on the observation of pair correlations, and bears some analogy with the study of dynamic deformation. 
The intersection with nuclear structure is outlined in Sec.~\ref{sec:structure}, focusing on the qualitative picture. 
Finally, in Sec.~\ref{sec:cumulants}, I introduce higher-order correlations, which are accurately measured in heavy-ion collisions, and explain why they are useful. 

The observables which have proven useful for the study of nuclear structure are the global ones, that involve all the particles seen in the detector. 
For this reason, I will only discuss the azimuthal anisotropy of an event as a whole. 
Note, however, that global observables are a small subset of the measured ones. 
Experiments also analyze in detail how anisotropies depends on transverse momentum, longitudinal momentum (rapidity), and particle species (meson versus baryon, strange and charmed hadrons, etc.). 

In addition to the multiplicity and the anisotropies which will be discussed below, there is one more relevant global observable, which this presentation does not cover, namely, the mean transverse momentum $\langle p_t\rangle$ of produced particles, whose importance for nuclear structure studies was first pointed out by Giacalone~\cite{Giacalone:2019pca}. 
I will only mention its relation with the temperature, which is essential. 
As we shall see shortly, the matter formed in a heavy-ion collision has a temperature $T$, which depends on position and time: 
The system cools down as a function of time, since it freely expands into the vacuum. 
At a given early time $t< R/c$ (such as depicted in the bottom panel of Fig.~\ref{fig:schema}), it is also hotter in the centre than near the edges, where the matter is less dense.
Despite this complexity, there is one relevant temperature scale, the effective temperature $T_{\rm eff}$, which corresponds roughly to the average over $(x,y)$ at $z=0$ and $t\simeq R/c$ (i.e., a time somewhat larger than that depicted in the bottom panel of Fig.~\ref{fig:schema}). 
It is related to the mean transverse momentum by $\langle p_t\rangle\simeq 3 T_{\rm eff}$~\cite{Gardim:2019xjs}.\footnote{In a grossly simplified picture, the system expands first longitudinally, then in the transverse directions. In this picture, $T_{\rm eff}$ corresponds to the temperature after the longitudinal expansion, and before the transverse expansion.}
A crucial observation in heavy-ion collisions is that $\langle p_t\rangle$ depends very weakly on impact parameter, and also of the size of the colliding nuclei at a given collision energy $\sqrt{s_{NN}}$~\cite{ALICE:2018hza}. 
The underlying reason is that the nuclear density is essentially universal, so that larger nuclei, or a more central collision, produces a larger system, but at the same effective temperature.  
This facilitates comparison between different nuclei, and is essential for nuclear structure studies, as will be detailed in Sec.~\ref{sec:structure}.

This article is dedicated to the memory of Art Poskanzer (1931-2021) who, in the later part of his career, was a key actor of this new field of research, of which he understood the importance from day one.   
He led the first analysis of elliptic flow by STAR which will be discussed in Sec.~\ref{sec:v2} and largely devised the method used to analyze it. 
He then played a crucial role in a number of subsequent developments, some of which I will mention below, and felt the significance of the results from U+U collisions at RHIC, which he encouraged me to study in detail in 2016, and which turned out to be the first link between nuclear structure and heavy ions, as I will discuss in Sec.~\ref{sec:structure}. 

\section{Elliptic flow in non-central collisions, and the old flow picture}
\label{sec:v2}

The RHIC collider started delivering Au+Au collisions at energy $\sqrt{s_{NN}}=130$~GeV in June 2000. 
Two results came out of the first day of data: 
The first result was the multiplicity of charged particles, measured by the PHOBOS collaboration~\cite{PHOBOS:2000wxz}, which theorists had been eagerly awaiting in order to compare it with their predictions.  
The second result, posted to arXiv by Art Poskanzer on Sept.12, 2000, was an unexpected breakthrough which had a long-lasting impact: 
The STAR collaboration observed a sizable azimuthal anisotropy called elliptic flow~\cite{STAR:2000ekf}, which had been predicted~\cite{Ollitrault:1992bk} and already observed in fixed-target experiments~\cite{E877:1994plr,NA49:1997qey}, but with a much smaller amplitude. 
(The history of these early measurements is sketched at the end of this section.)

\begin{figure}[tb]
\begin{center}
\includegraphics[width=.8\linewidth]{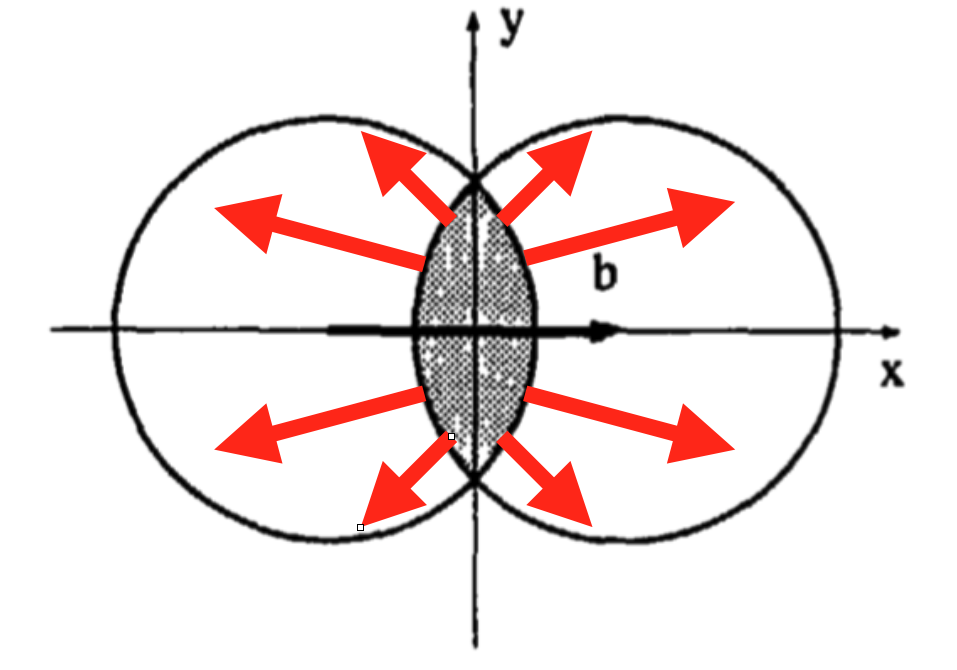}
\end{center}
\caption{Schematic view of a collision between two identical spherical nuclei at impact parameter $b$, in the transverse plane $z=0$~\cite{Ollitrault:1993iw}.
Matter is produced in the shaded overlap area, where the participant nucleons lie. 
Red arrows represent pressure gradients within the produced matter. 
Outgoing particles typically go in the same directions as pressure gradients. 
This results in an elliptic anisotropy of their distribution.      
}
\label{fig:b}
\end{figure}

Elliptic flow is best seen in collisions with a large impact parameter, as depicted in Fig.~\ref{fig:b}.  
I first explain the mechanism. 
As explained above, matter is produced where the nuclei overlap. 
Spectator nucleons, which are outside the overlap area, continue their motion along the beam direction and quickly disappear from the interaction area. 
What is left is a blob of quark-gluon matter  with the shape of an almond, which is globally at rest in the transverse plane, since the initial motion of the nuclei is purely longitudinal. 
Strong interactions among quarks and gluons generate pressure, which is largest at the center of the almond, and gradually decreases to zero as one moves towards the edges. 
It generates a pressure-gradient force, directed toward regions of lower pressure, which pushes the matter outward in all directions. 
The force density has a component $f_x=-\partial P/\partial x$ along $x$, and $f_y=-\partial P/\partial y$ along $y$. 
Now, the size of the almond is smaller along the $x$ direction, which implies on average $\left|f_x\right|>\left|f_y\right|$: 
The force field is anisotropic, as depicted by the red arrows in Fig.~\ref{fig:b}. 
The expansion generated by pressure gradients can be computed by modeling the produced matter as an ideal relativistic fluid~\cite{Ollitrault:1992bk,Kolb:2000sd},\footnote{This description was later refined by taking viscosity into account~\cite{Romatschke:2007mq}. Note also that back in 2000, the pressure of QCD was largely unknown. It has since then been accurately computed as a function of temperature~\cite{Borsanyi:2013bia}, and all modern fluid-dynamical calculations use this equation of state as input~\cite{JETSCAPE:2020mzn}.} which eventually fragments into individual particles (``freeze out''). 
The directions of particles seen in the detector retain the memory of the anisotropic force field. 
Decomposing the velocity of a particle as $v_x=v\cos\phi$, $v_y=v\sin\phi$, where $\phi$ is the angle between the $x$ axis and the velocity, one therefore expects on average $\left|\cos\phi\right|>\left|\sin\phi\right|$. 
Squaring and taking the difference, one expects $\cos2\phi=\cos^2\phi-\sin^2\phi$ to be positive on average. 

\begin{figure}[tb]
\begin{center}
\includegraphics[width=\linewidth]{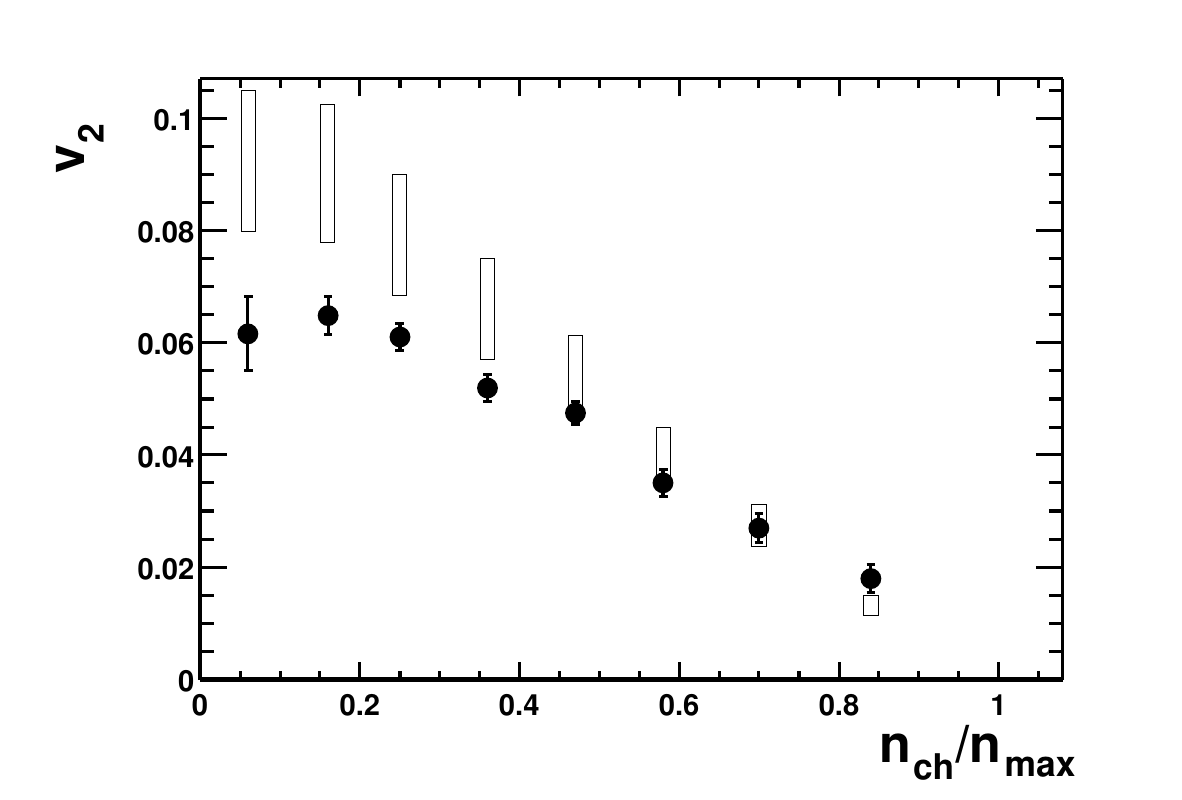}
\end{center}
\caption{Discovery of elliptic flow by the STAR collaboration in Au+Au collisions at $\sqrt{s_{NN}}=130$~GeV~\cite{STAR:2000ekf}. 
Events are sorted according to the particle multiplicity $n_{ch}$ shown on the $x$ axis. 
It is scaled by $n_{max}$, defined as is the largest value of $n_{ch}$ in the analyzed set of collision events.  
Boxes represent early hydrodynamic predictions~\cite{Kolb:2000sd}, based on ideal fluid dynamics. 
}
\label{fig:STAR}
\end{figure}

In the ``old flow picture'', which will be defined precisely below, elliptic flow, denoted by $v_2$~\cite{Voloshin:1994mz}, is defined as the average value of $\cos2\phi$ over all particles and all events with the same $b$. 
It lies between $-1$ and $+1$, where the extreme values correspond to the cases where all trajectories are parallel to $y$ ($\phi=\frac{\pi}{2}$) or to $x$ ($\phi=0$). 
One expects in general $v_2>0$. 
This phenomenon is also referred to as ``in-plane'' elliptic flow, in  reference to the reaction plane $(x,z)$, defined as the plane containing the collision axis and the impact parameter. 
Note that the average value of $\sin 2\phi$ is zero, because the geometry of Fig.~\ref{fig:b} is symmetric under the transformation $y\to -y$, that is, $\phi\to-\phi$, and parity conservation requires that this symmetry is preserved by the collision dynamics.

Fig.~\ref{fig:STAR} displays the variation of $v_2$ with the centrality of the collision which, as explained above, is evaluated from the number of particles seen in the detector (it only detects charged particles, hence the notation $n_{ch}$ on the $x$ axis): 
 $b$ decreases as one moves from the left to the right of the diagram. 
 $v_2$ is positive\footnote{The sign of $v_2$ was actually not measured in this first analysis, as will be explained below.} for all centralities. 
It is large in collisions with large impact parameter, where the almond shape is very anisotropic, and decreases gradually as the collision becomes more central, as expected from the qualitative picture outlined above. 

The big surprise from these first RHIC data was that $v_2$ was almost as large as predictions from ideal-fluid models, showed as boxes in Fig.~\ref{fig:STAR}, which are believed to represent an upper bound on the effect.  
This gradually led to the ``perfect liquid'' picture~\cite{Jacak:2010zz}, in which the collision creates a strongly-interacting quark-gluon plasma. 
Hydrodynamic models have been much refined since then, and many more data have been collected, but the main conclusion still remains. 
More specifically, global theory-to-data comparisons~\cite{JETSCAPE:2020mzn}  favor values of the viscosity such that $v_2$ is smaller only by  $\sim 20\%$ than the ``ideal hydrodynamic limit'' in Pb+Pb collisions at the LHC~\cite{Gardim:2022vys}, but the uncertainty on the viscosity remains large, and a larger viscous suppression (say, $30\%$) is not excluded. 

\begin{figure}[tb]
\begin{center}
\includegraphics[width=\linewidth]{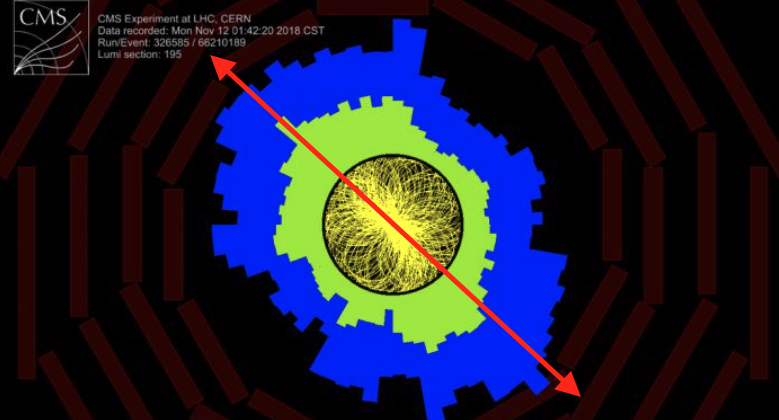}
\end{center}
\caption{Event display of a non-central Pb+Pb collision at $\sqrt{s_{NN}}=5.02$~TeV in the transverse plane, seen by the CMS detector. 
The yellow lines are the charged-particle trajectories, and the green and blue shaded areas show the energy deposited in the electromagnetic and hadronic calorimeters, respectively~\cite{Velkovska}.
The red arrow indicates the probable direction of impact parameter, as inferred from the elliptical shape of the event. 
}
\label{fig:CMS}
\end{figure}

The RHIC result was inferred from a statistical analysis carried out on a large number of events, as I will explain shortly.
In a single event, the elliptic anisotropy is not clearly visible, because the number of particles seen in the detector is not large enough. 
By contrast, at the LHC, one typically sees elliptic flow by eye in event displays, as illustrated in Fig.~\ref{fig:CMS}. 
The direction where more particles are emitted is that of impact parameter, so that one can immediately guess the orientation of the reaction plane. 
The difference between RHIC and LHC is the result of two factors. 
A larger number of particles is produced, due to the higher energy, and a larger fraction of these particles are detected, due to better detector acceptance: 
The CMS and ATLAS detectors at the LHC, in particular, cover a larger fraction of the solid angle around the collision point.  

I now discuss how the value of $v_2$ is measured. 
If the reaction plane was known in every event, this would be a very easy measurement. One would define that direction as the $x$ axis, as in Fig.~\ref{fig:b}, and evaluate the average value of $\cos 2\phi$ over all particles from all events in a centrality class. 
But the reaction plane is not known and varies event to event.
There is no preferred direction because the size of nuclei is smaller by many orders of magnitude than the beam diameter, so that the probability distribution of the reaction plane is uniform.  

\begin{figure}[tb]
\begin{center}
\includegraphics[width=.9\linewidth]{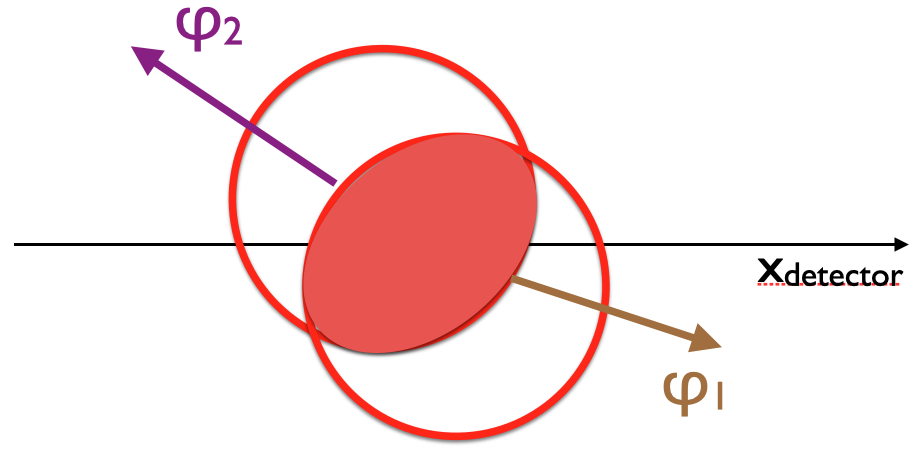}
\end{center}
\caption{Schematic view of a symmetric nucleus-nucleus collision in the transverse plane $z=0$, as seen in the laboratory frame. 
The orientation of the almond is random, as opposed to the theorist view of Fig.~\ref{fig:b} where it is aligned along a well-defined axis $x$.  
Elliptic flow is measured by correlating pairs of particles with azimuthal angles $\phi_1$ and $\phi_2$ (see text). 
}
\label{fig:randomb}
\end{figure}

The only information is the direction of outgoing particles,\footnote{Equivalent information is obtained in calorimeters or scintillators, which record an amplitude as a function of azimuthal angle $\phi$.} whose azimuthal angles $\phi$ are measured with respect to a fixed direction in the detector (Fig.~\ref{fig:randomb}), and $v_2$ must be expressed as a function of these azimuthal angles.
The simplest method is a two-particle correlation~\cite{Wang:1991qh}. 
One takes a pair of particles belonging to the same event, with angles $\phi_1$ and $\phi_2$ (Fig.~\ref{fig:randomb}). 
Since these angles are measured with respect to an arbitrary reference direction in the detector, the only relevant quantity is the difference $\Delta\phi\equiv\phi_1-\phi_2$, which is independent of the reference. 

One then constructs the following correlator:
\begin{equation}
  \label{defc22}
c_2\{2\}\equiv \left\langle e^{2i(\phi_1-\phi_2)}\right\rangle,
\end{equation}
where the subscript in $c_2$ refers to the factor 2 in front of $(\phi_1-\phi_2)$, and the notation $\{2\}$ refers to the number of particles which are correlated~\cite{Borghini:2001vi}, which can be larger than $2$ as will be discussed in Sec.~\ref{sec:cumulants}. 
The angular brackets in the right-hand side denote an average over pairs of particles of all events.
If the $i^{\rm th}$ event has $N_i$ particles, then it has $N_i(N_i-1)/2$ pairs, and the number of terms in the average is $\sum_i N_i(N_i-1)/2$, where the sum runs over events~\cite{Bilandzic:2010jr}.\footnote{In practice, some of the pairs are excluded, as will be explained in detail in Sec.~\ref{sec:ridge}.}

The next step is the most important one, which is usually referred to as the ``flow paradigm''. 
One assumes that particles are emitted independently in the ``intrinsic frame'' of Fig.~\ref{fig:b}, where the direction of impact parameter is fixed. 
The motivation for this assumption is that if that interactions in the quark-gluon matter are strong enough, the system thermalizes (which is an underlying assumption of fluid dynamics) and correlations are lost. 
With help of this hypothesis, one can easily relate the measured correlator $c_2\{2\}$ to $v_2$. 
If we express the angles $\phi_1$ and $\phi_2$ in the intrinsic frame, independence implies that the expectation value of the product is a product of expectation values: 
\begin{equation}
  \label{factorization}
  \left\langle e^{2i(\phi_1-\phi_2)}\right\rangle=\left\langle e^{2i\phi_1}\right\rangle\left\langle e^{-2i\phi_2}\right\rangle=(v_2)^2,
\end{equation}
where, in the last equality, we have used $\langle\sin 2\phi\rangle=0$ and $\langle\cos 2\phi\rangle=v_2$.
Thus, $v_2$ is obtained by taking the square root of $c_2\{2\}$.
This estimate of $v_2$ is denoted by $v_2\{2\}$:
\begin{equation}
  \label{defv22}
v_2\{2\}\equiv \left(c_2\{2\}\right)^{1/2}. 
\end{equation}

Note that the pair correlation gives access to $(v_2)^2$, rather than $v_2$, and that the sign of $v_2$ cannot be determined with this method. 
Experimental evidence that $v_2$ is positive at RHIC was obtained a few years later using a different method, in a letter posted by Art Poskanzer on behalf of the STAR Collaboration~\cite{STAR:2003xyj}. 
But already in 2000, $v_2$ was {\it thought\/} to be positive. 
The sign was no longer controversial, but had been the subject of intense discussions in the last decade of the twentieth century~\cite{Ollitrault:1997vz}, which I briefly recall. 
The first heavy-ion collisions were done at much lower energies, where the Lorentz contraction is modest. 
What happens then is that spectator nucleons stay in the way long enough that they block emission of particles in the reaction plane~\cite{Ollitrault:1993ba}, resulting in out-of-plane emission, that is, negative $v_2$. 
This phenomenon, called ``squeeze-out'', was observed for nucleons in 1990~\cite{Gutbrod:1989gh}, then for pions~\cite{Brill:1993xh}. 
It was so well established that the first claim that $v_2$ was positive at higher energies, in 1994~\cite{E877:1994plr}, first raised some doubts. 
Conclusive evidence only came in 1997, first at the AGS in Brookhaven~\cite{E877:1997zjw} and shortly after at the SPS at CERN~\cite{NA49:1997qey} (yet another letter posted by Art Poskanzer). 
The energy at which $v_2$ changes sign from negative to positive was determined in 1999~\cite{E895:1999ldn}, one year before RHIC began operation. 

For the sake of completeness, let me finally mention that the method used by STAR to obtain the results in Fig.~\ref{fig:STAR} was not, strictly speaking, a pair correlation. 
The analysis was done with the event-plane method, established by Poskanzer and Voloshin~\cite{Poskanzer:1998yz,Ollitrault:1997di}. 
In this method, which is inspired by collisions at lower energies~\cite{Danielewicz:1985hn}, every particle is correlated with an estimate of the reaction plane (called the ``event plane'') constructed by using many other particles. 
This technique was later shown~\cite{Ollitrault:2009ie} to be essentially equivalent to a simple pair correlation~\cite{Wang:1991qh}, and actually worse when it comes to precision measurements~\cite{Luzum:2012da}, so that it is less used in recent analyses (the original title of Ref.~\cite{Luzum:2012da}, which was  changed by the editors of Phys. Rev. C., was  ``The event-plane method is obsolete''). 

In deriving Eq.~(\ref{factorization}), we have implicitly assumed that the equations $\langle\sin 2\phi\rangle=0$ and $\langle\cos 2\phi\rangle=v_2$ hold for every event. 
That is, we have postulated that all collisions with the same impact parameter are identical, up to statistical fluctuations. 
This assumption (which was always implicit in the first flow papers) defines the ``old flow picture''. 

\section{Modern picture of azimuthal anisotropy}
\label{sec:ridge}

The old flow picture was gradually replaced by a more modern picture, in which events with the same impact parameter are no longer assumed to be identical up to statistical fluctuations. 
Event-by-event fluctuations were modeled in hydrodynamic calculations as early as 2000 by the Brazilian group~\cite{Aguiar:2001ac}, but their relevance for understanding experimental data from RHIC was recognized only a decade later. 

\subsection{Two puzzles}
\label{sec:puzzles}

Two puzzling results came out of RHIC experiments around the year 2005.\footnote{I borrow some of the following words from the introduction of the PhD thesis of Burak Alver~\cite{alverthesis} who, together with his supervisor Gunther Roland, gets the credit for the modern picture of azimuthal anisotropy.}
The first puzzle was the observation of sizable elliptic flow even in central Au+Au collisions, where the initial geometry is expected to be roughly circular. 
One sees in Fig.~\ref{fig:STAR} that the rightmost data point is above the theoretical prediction, while all other points are below. 
This was considered a detail until the analysis was repeated with smaller nuclei, in Cu+Cu collisions. 
Not only $v_2$ was non zero in central Cu+Cu collisions, it was even {\it larger\/} than in central Au+Au collisions~\cite{PHOBOS:2005gex,PHOBOS:2006dbo}!

The second puzzling observation was a peculiar structure seen in pair correlations, which I will discuss in detail as I consider it the most spectacular observation in heavy-ion collisions. It was first observed at RHIC~\cite{STAR:2004wfp}, but I will show instead more recent data from the LHC~\cite{CMS:2012xss} which are similar and of better quality, so that the structure appears more clearly. 

\begin{figure}[tb]
\begin{center}
\includegraphics[width=\linewidth]{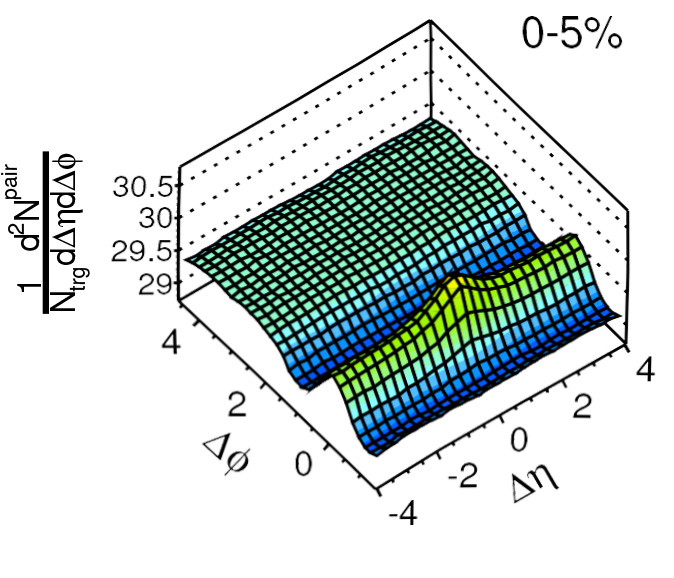}
\end{center}
\caption{Number of particle pairs as a function of relative azimuthal angle $\Delta\phi\equiv\phi_1-\phi_2$ and pseudorapidity $\Delta\eta\equiv \eta_1-\eta_2$, in central Pb+Pb collisions at $\sqrt{s_{NN}}=2.76$~TeV \cite{CMS:2012xss}. 
$\Delta\eta$ corresponds physically to the relative longitudinal velocity of the pair (see text). 
}
\label{fig:cmsridge}
\end{figure}

The idea is to study the distribution of the relative direction between two particles, in a more detailed way than the $c_2\{2\}$ defined in Sec.~\ref{sec:v2}. 
Each particle has, in addition to its azimuthal angle $\phi$, a polar angle $\theta$ with respect to the collision axis. 
I first explain how the polar angle is used to reconstruct the relative longitudinal velocity. 
Particles used in the analysis have velocities close to $c$\footnote{The data shown in Fig.~\ref{fig:cmsridge} correspond to specific cuts on the transverse momenta of the particles ($3<p_t<3.5$~GeV$/c$ for one particle and  $1<p_t<1.5$~GeV$/c$ for the other). The vast majority of particles are pions with mass energy $mc^2\simeq 0.14$~GeV. Momenta are much larger than $mc$, which implies that velocities are very close to $c$. } so that the longitudinal velocity is $v_z\simeq c\cos\theta$. 
In special relativity, however, the relative velocity between two particles (the velocity of one in the rest frame of the other) is not the difference of their velocities. 
Therefore, one prefers an equivalent variable, the rapidity $y$, for which additivity is restored. 
It is defined by $\tanh y\equiv v_z/c$. 
Since $v_z/c$ is almost equal to $\cos\theta$, the {\it pseudo}rapidity $\eta$, defined by $\tanh\eta=\cos\theta$, is almost equal to the rapidity $y$. 
Hence, the relative pseudorapidity $\Delta\eta$ between two particles measures their relative longitudinal velocity. 

Fig.~\ref{fig:cmsridge} displays the number of pairs as a function of the relative azimuthal angle $\Delta\phi$ and the relative pseudorapidity $\Delta\eta$. 
The most remarkable feature is that apart from a small peak around the origin $\Delta\phi=\Delta\eta=0$, it is essentially independent of $\Delta\eta$. 
This feature is specific to nucleus-nucleus collisions, and it not at all observed in proton-proton collisions. 
It is a typical emergent phenomenon, where the larger complexity, due to the larger number of particles, eventually results in a simpler picture.\footnote{As argued by Anderson~\cite{Anderson:1972pca}, emergent phenomena are usually associated with broken symmetries. In the case of Fig.~\ref{fig:cmsridge}, we will see that the origin of the observed pattern lies in the breaking of azimuthal symmetry in every event.}
The rich structure lies in the dependence on $\Delta\phi$. 
The maximum around $\Delta\phi=0$ is referred to as the ``near-side ridge''.\footnote{This term was coined  by Joern Putschke at the Quark Matter 2006 conference~\cite{Putschke:2007mi} because the structure, with a long ridge surrounding a small peak, reminded him of the Weisshorn peak in the Alps, which is climbed via ridges~\cite{Putschkeridge}.}. 
This ridge is surrounded by two symmetric minima. 
As $\Delta\phi$ further increases, the number of pairs increases again and reaches a wide plateau around $\Delta\phi=\pi$, called the ``away side'' because it corresponds to pairs emitted back-to-back in azimuth. 
None of these structures was understood at first. 
I will first explain how we now understand them, and then briefly sketch the path that led to this understanding, which is instructive from a historical point of view. 

\subsection{Mathematical solution}
\label{sec:paradigm}

The solution to both puzzles is deceptively simple. 
It was understood in 2010 by Alver and Roland~\cite{Alver:2010gr} and clearly formulated by Luzum one year later~\cite{Luzum:2011mm}. 
I will first explain the mathematics. 
The argument consists of two parts. 
First, one assumes that the directions $(\eta,\phi)$ of particles in a given collision event are sampled independently from an underlying probability distribution $P(\eta,\phi)$~\cite{Luzum:2011mm}. 
That is, particles in a single event are uncorrelated, according to the ``flow paradigm'' already mentioned above. 
Second, one assumes that this distribution is independent of $\eta$, so that it can just be written as $P(\phi)$. 

 $P(\phi)$ can be any $2\pi$-periodic function. 
One decomposes it as a Fourier series: 
\begin{equation}
\label{fourier}
P(\phi)=\frac{1}{2\pi}\sum_{n=-\infty}^{+\infty}V_ne^{-in\phi}, 
\end{equation}
where $V_n$ is a complex Fourier coefficient,\footnote{The notation $v_n$ for Fourier coefficients was introduced in the context of the old flow picture by Voloshin and Zhang~\cite{Voloshin:1994mz}.} which is obtained from $P(\phi)$ through the standard inversion formula: 
\begin{equation}
\label{inversefourier}
V_n= \int_0^{2\pi}e^{in\phi}P(\phi)d\phi. 
\end{equation}
Note that $V_0=1$ because probabilities must add up to unity, $\int_0^{2\pi}P(\phi)d\phi=1$. 
Other Fourier coefficients vary event to event, both in phase and magnitude. 
Since $P(\phi)$ is real, they satisfy $V_{-n}=V_n^*$. 
The first harmonics, $n=1,2,3,4$, are called directed~\cite{Ollitrault:1997vz}, elliptic, triangular~\cite{Alver:2010gr}, and quadrangular~\cite{Giacalone:2016mdr} flow. 
 
Let us evaluate the pair distribution in this simple model. 
In a single event, independence implies that it is the product of single-particle distributions:
\begin{equation}
\label{pair}
\frac{dN_{\rm pair}}{d\eta_1d\phi_1d\eta_2d\phi_2}=P(\phi_1)P(\phi_2),
\end{equation}
up to a normalization constant, which will be omitted also in the following equations.\footnote{
In writing this equation, I have assumed for simplicity that the single-particle distribution (\ref{fourier}) is the same for both particles in the pair. 
It is not the case for the CMS analysis in Fig.~\ref{fig:cmsridge} since the two particles are chosen in separate $p_t$ intervals, but this is a detail which does not alter the general picture.} 
Writing $\phi_1=\Delta\phi+\phi_2$, substituting (\ref{fourier}) and integrating over $\phi_2$, one obtains:
\begin{eqnarray}
\label{fourierpair}
\frac{dN_{\rm pair}}{d\eta_1d\eta_2d\Delta\phi}&=&\frac{1}{2\pi}\sum_{n=-\infty}^{+\infty}V_nV_{-n}e^{-in\Delta\phi}
\cr
&=&\frac{1}{2\pi}\left(	1+2  \sum_{n=1}^{+\infty}\left|V_n\right|^2 \cos(n\Delta\phi)\right).
\end{eqnarray}
This equation is for a single event, and must then be averaged over events: 
\begin{equation}
\label{fourierpairav}
\left\langle\frac{dN_{\rm pair}}{d\eta_1d\eta_2d\Delta\phi}\right\rangle=\frac{1}{2\pi}\left(	1+2  \sum_{n=1}^{+\infty}\left\langle \left|V_n\right|^2\right\rangle\cos(n\Delta\phi)\right),
\end{equation}
where angular brackets denote an average over events in a centrality class. 
Since we have assumed that the single-particle distribution is independent of $\eta$, the pair distribution is independent of  $\Delta\eta=\eta_1-\eta_2$. 

The remarkable feature of Eq.~(\ref{fourierpairav}) is that the Fourier coefficients of the distribution of $\Delta\phi$, namely,  $\left\langle \left|V_n\right|^2\right\rangle$, are all positive. 
This implies that the absolute maximum of the pair distribution is at $\Delta\phi=0$, where all the cosines reach their maximum value $1$, in agreement with the observation in  Fig.~\ref{fig:cmsridge}.  
The flow paradigm alone implies that there is a near-side ridge, and that it is higher than the away-side, even without specifying the azimuthal distribution $P(\phi)$. 
The specific structure seen in the figure, with a very flat plateau on the away side, is explained by keeping only two harmonics in the sum, $V_2$ and $V_3$, with $\left|V_3\right|\simeq \frac{2}{3}\left|V_2\right|$.  
The breakthrough of Alver and Roland, in 2010, was to propose triangular flow, $V_3$, as an explanation for the structure seen in pair correlations~\cite{Alver:2010gr}. 

To summarize, the solution to the puzzle is that particles are emitted independently in every event. 
It may seem a paradox that {\it independent\/} particle emission explains data on {\it correlations\/}. 
The key is that in a single event,  the single-particle distribution $P(\phi)$ depends on $\phi$, so that azimuthal anisotropy is broken, in a way that  {\it fluctuates\/} event to event. 
An alternative way of formulating the flow paradigm is to say that all correlations arise from event-by-event fluctuations. 
As we shall see in Sec.~\ref{sec:cumulants}, such fluctuations generate correlations to all orders, not just pair correlations. 

What the flow paradigm cannot explain is the small peak around $\Delta\phi=\Delta\eta=0$, corresponding to particles emitted at small angle with respect to one another.\footnote{The angle between the particles is $\left(\Delta\phi^2+\Delta\eta^2\right)^{1/2}$ when $\Delta\phi$ and $\Delta\eta$ are both much smaller than unity.}
This peak must be attributed to correlations between outgoing particles, which are generically referred to as ``nonflow'' because they fall out of the flow paradigm. 
In the case of Fig.~\ref{fig:cmsridge}, where one of the particles has fairly large transverse momentum, the correlation is believed to result from the formation of a jet or ``minijet'', defined as the emission of a collimated shower of particles. 

\subsection{Physical mechanism}

\begin{figure}[tb]
\begin{center}
\includegraphics[width=.8\linewidth]{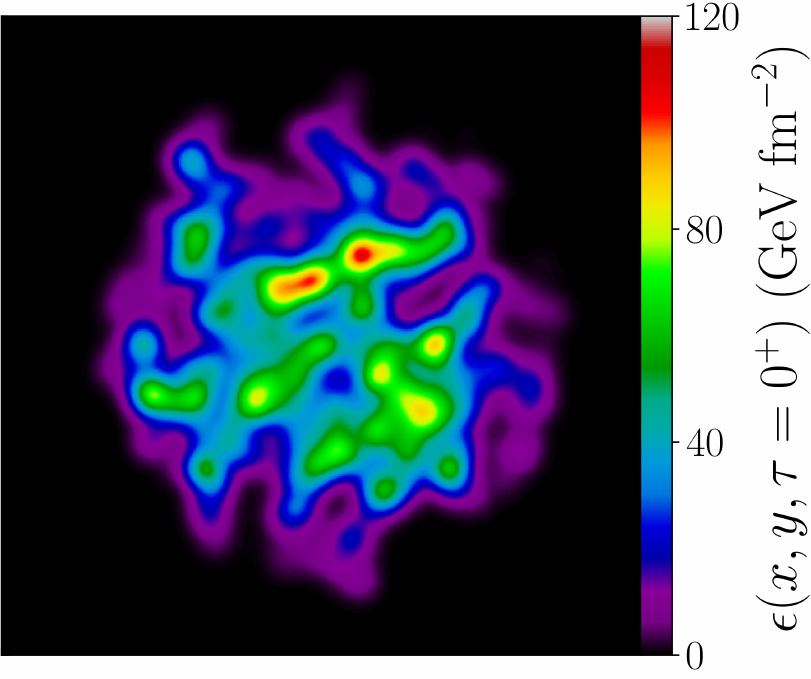}
\end{center}
\caption{Model calculation of the distribution of energy in the transverse plane, $\epsilon(x,y)$,  at early times in a random central Pb+Pb collision at the LHC~\cite{Giacalone:2022hnz}.
}
\label{fig:profile}
\end{figure}
Let me now sketch the physical picture behind the flow paradigm. 
As explained in Sec.~\ref{sec:v2}, the assumption of independent particles is a natural consequence of the formation of a fluid, which implies an underlying thermalization mechanism washing out correlations. 
Now, the equations of fluid dynamics are deterministic. 
The single-particle probability distribution $P(\eta,\phi)$, which determines the emission of particles, is uniquely determined by their initial conditions. 
These initial conditions consist of the density profile of matter created very shortly after $t=0$. 
As explained above, the collision captures a specific quantum state of the incoming nuclei. 
The thickness functions of incoming nuclei, $T_A(x,y)$ and $T_B(x,y)$, are distributed irregularly, depending on the positions of nucleons at the time of impact~\cite{PHOBOS:2005gex}. 
These fluctuations are reflected in the density of produced matter, as illustrated in Fig.~\ref{fig:profile}. 
Despite large theoretical uncertainties~\cite{Giacalone:2022hnz,Nijs:2023yab}, there is general consensus that such fluctuations are present, and that they are approximately independent of the longitudinal coordinate $z$. 

The symmetry of the fluid is carried over to the distribution of particles it emits. 
The longitudinal velocity is approximately given by $v_z=z/t$~\cite{Bjorken:1982qr}, so that if fluctuations are independent of $z$, they are independent of the longitudinal velocity, which implies in turn that the probability $P(\eta,\phi)$ is approximately independent of $\eta$. 
On the other hand, it can be any function of $\phi$. 
Due to fluctuations, central collisions are no longer azimuthally symmetric, and $V_2$ can differ from zero, which solves the first puzzle of  Sec.~\ref{sec:puzzles}. 
$V_2$ can also have an imaginary part, generated by a non-zero $\langle\sin 2\phi\rangle$ in the intrinsic frame~\cite{PHOBOS:2006dbo}. 

\subsection{Fourier spectrum of azimuthal anisotropies}
\label{sec:vn2}

 \begin{figure}[ht]
    \centering
\includegraphics[width=\linewidth]{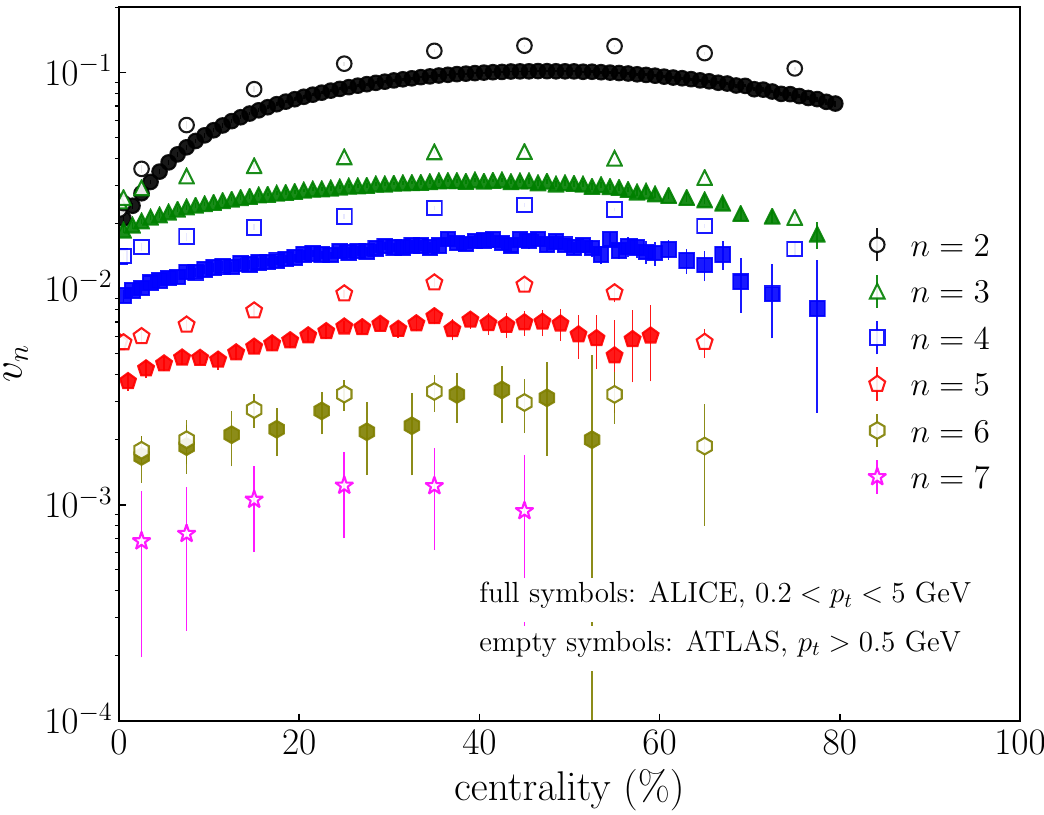}
    \caption{Centrality dependence of $v_n\{2\}$, with $n=2,\cdots,7$, in Pb+Pb collisions at $\sqrt{s_{\rm NN}}=5.02$~TeV. 
 ALICE data~\cite{Acharya:2018lmh} are shown as full symbols and ATLAS data~\cite{Aaboud:2018ves} as empty symbols.
  $v_7\{2\}$ data are from ATLAS only (Courtesy of Giuliano Giacalone).}
    \label{fig:vncentrality}
\end{figure}

Experimentally, one measures the Fourier coefficients of the distribution of $\Delta\phi$, defined through an immediate generalization of Eq.~(\ref{defc22}): 
\begin{equation}
  \label{defcn2}
c_n\{2\}\equiv \left\langle e^{in\Delta\phi}\right\rangle=\left\langle \left| V_n\right|^2\right\rangle, 
\end{equation}
where, in the last equality, we have used Eq.~(\ref{fourierpairav}). 
In practice, the average is not done over all pairs, but only over those such that $|\Delta\eta|$ is larger than a gap which depends on the analysis: ALICE data shown in Fig.~\ref{fig:vncentrality} are obtained with $|\Delta\eta|>1$, and ATLAS data with $|\Delta\eta|>2$.
This rapidity gap is meant to eliminate the nonflow peak near $\Delta\phi=\Delta\eta=0$, and isolate the long-range correlation (long range meaning large $|\Delta\eta|$), which is thought to be dominated by independent particle emission. 

As in Eq.~(\ref{defv22}), $v_n\{2\}$ is defined by taking  the square root of $c_n\{2\}$; 
 \begin{equation}
  \label{defvn2}
v_n\{2\}\equiv \left(c_n\{2\}\right)^{1/2}=\left\langle \left| V_n\right|^2\right\rangle^{1/2}.
\end{equation}
$V_n$ fluctuates event to event, and $v_n\{2\}$ is its rms (root mean square) value. 
Fig.~\ref{fig:vncentrality} displays the value of $v_n\{2\}$, for $n=2,\dots,7$ in  Pb+Pb collisions at $\sqrt{s_{\rm NN}}=5.02$~TeV as a function of centrality percentile. 
ALICE and ATLAS differ because of a different acceptance. 
ATLAS does not detect particles with low transverse momenta $p_t$, which spiral without reaching the inner detector due to a large magnetic field. 
Since the azimuthal anisotropy typically increases with $p_t$, the rms average is larger when low $p_t$ particles are excluded.  
This explains why ATLAS data points are systematically above ALICE data points.   

There is a strong hierarchy among harmonics of different order. 
The Fourier spectrum is dominated by elliptic flow. 
As will be shown in Sec.~\ref{sec:cumulants}, the static (in the sense that it does not fluctuate event to event) elliptic deformation discussed in Sec.~\ref{sec:v2} is the leading contribution to $v_2\{2\}$, except for the most central collisions. 
In the most central collisions, $v_2\{2\}$ and $v_3\{2\}$ are of the same magnitude, which is not yet fully understood and referred to as the ``ultracentral puzzle''~\cite{Luzum:2012wu,Giannini:2022bkn}.\footnote{We have pointed out that CMS results in Fig.~\ref{fig:cmsridge} are compatible with a $v_3$  close to $\frac{2}{3}v_2$. This is compatible with the findings of ALICE and ATLAS, if one averages them over the $0-5\%$ centrality range used by CMS.}
The possible role of an octupole deformation of $^{208}$Pb, which would enhance $v_3\{2\}$ as will be discussed in Sec.~\ref{sec:structure}, has been investigated~\cite{Carzon:2020xwp}. 

Note also that the first harmonic $n=1$, corresponding to directed flow, is not shown. 
It is typically smaller than triangular flow, but of comparable magnitude~\cite{Teaney:2010vd}. 
It is more difficult to measure for two reasons, which both have to do with the condition that the total transverse momentum of every event must be $0$, as a consequence of momentum conservation. 
First, this condition implies that in every event,  the real and imaginary parts of $V_1$ both change sign as a function of $p_t$~\cite{Teaney:2010vd}, so that results depend even more strongly on the $p_t$ range chosen for the analysis. 
Second, the condition that the transverse momenta of particles sum up to zero in every event induces a nonflow correlation which is specific to $v_1$~\cite{Borghini:2000cm} and must be carefully subtracted~\cite{Retinskaya:2012ky,ATLAS:2012at}. 
Because of these difficulties, directed flow is much less studied than higher harmonics. 

\subsection{A bit of history}

Despite the simplicity of the modern flow picture, it took years to emerge. 
The first puzzle, that of large elliptic flow in central collisions, was solved rather quickly. 
Miller and Snellings pointed out already in 2003 that the shape of the almond may fluctuate event by event~\cite{Miller:2003kd}. 
The PHOBOS collaboration, who analyzed the Cu+Cu data, pointed out that the orientation of the almond (corresponding to the phase of $V_2$)  can also fluctuate event to event, and that these fluctuations explain observations. 
This was a highlight of the Quark Matter 2005 conference~\cite{PHOBOS:2005gex,PHOBOS:2006dbo}. 

The second puzzle, the near-side ridge and the away-side plateau, required a lot more work, and was the subject of intense activity in the years 2005-2010. 
The title of the first paper where this phenomenon was observed~\cite{STAR:2004wfp} started with ``minijet deformation'', making it sound like the physics had to do with jets, which may have misled theorists. 
It was first thought that the near-side ridge resulted from a broadening of the jet peak along the $\Delta\eta$ direction, and it was argued that the energy loss of a jet traversing a quark-gluon plasma (jet quenching) would produce such an effect~\cite{Majumder:2006wi,Dumitru:2007rp}. 
Similarly, the away-side plateau was first thought to result from a modification of the back jet (jets usually go in pairs, emitted back to back). 
The proposed mechanism was that a supersonic jet traversing a quark-gluon plasma would emit a Mach cone~\cite{Casalderrey-Solana:2004fdk,Satarov:2005mv,Ruppert:2005uz,Koch:2005sx}, and the away-side plateau was often referred to as a ``conical structure''. 

It was then realized that a near-side ridge can be generated by longitudinally-extended hot spots, boosted by the transverse expansion~\cite{Shuryak:2007fu,Dumitru:2008wn,Gavin:2008ev}. 
This picture is consistent with the modern understanding, but less simple. 
Event-by-event hydrodynamic calculations by the Brazilian group soon managed to quantitatively reproduce the data~\cite{Takahashi:2009na}, which was a clear hint that  flow played a crucial role. 
The state of confusion of the field was summarized by Nagle at the Quark Matter 2009 conference~\cite{Nagle:2009wr}. 
The breakthrough of Alver and Roland, in early 2010~\cite{Alver:2010gr}, was to realize that triangular flow explained simultaneously the near-side ridge and the away-side plateau. 
More hydrodynamic calculations soon followed and confirmed this explanation~\cite{Alver:2010dn,Petersen:2010cw,Schenke:2010rr}. 
It was the final proof that the observed structure had nothing to do with jets, but was all generated by flow. 

The same year, the near-side ridge acquired fame beyond the heavy-ion community, when it was observed in high-multiplicity proton-proton collisions at the LHC~\cite{CMS:2010ifv}, where nobody expected it. 
It was the first significant result from the LHC, and this phenomenon has challenged the best established models of proton-proton collisions~\cite{Skands:2014pea}.  

\section{Relating azimuthal anisotropy with nuclear structure}
\label{sec:structure}

\subsection{Anisotropic flow is generated by initial anisotropies}
\label{sec:epsn}

I now briefly explain why the azimuthal anisotropies measured in heavy-ion collisions are relevant to nuclear structure studies~\cite{Giacalone:2020ymy}. 
The main point is that the azimuthal anisotropy of outgoing particles is closely related to that of the initial energy density profile created just after the collision. 
In the same way as we have argued, in Sec~\ref{sec:v2}, that an overlap area with an almond shape, similar to an ellipse, generates elliptic flow, a triangular overlap area would generate triangular flow~\cite{Alver:2010gr}, simply because the $\phi\to\phi+\frac{2\pi}{3}$ symmetry is preserved throughout the evolution. 

Now, a fluctuating density profile such as depicted in Fig.~\ref{fig:profile} is neither elliptic nor triangular. 
However, one can carry out a Fourier decomposition of $\epsilon(x,y)$ at early times~\cite{Teaney:2010vd} and isolate its ``ellipticity'' $\varepsilon_2$ and its ``triangularity'' $\varepsilon_3$. 
In the center of mass of the energy, they are defined in complex form by~\cite{Qiu:2011iv}:
\begin{equation}
\varepsilon_n\equiv \frac{\int r^n e^{in\phi}\epsilon(x,y)dxdy}{\int r^n \epsilon(x,y)dxdy}, 
\end{equation}
where $(r,\phi)$ are the usual polar coordinates in the $(x,y)$ plane. 
The factor $r^n$ in the numerator ensures analyticity of the integrand, since $ r^n e^{in\phi}=(x+iy)^n$. 
The factor $r^n$ in the denominator, on the other hand, makes $\varepsilon_n$ dimensionless and contained within the unit circle,  $\left|\varepsilon_n\right|<1$, much like an actual Fourier coefficient. 

Numerical hydrodynamic calculations show that, to a good approximation, $V_2$ and $V_3$ are determined by linear response to the corresponding initial anisotropies~\cite{Niemi:2012aj,Gardim:2014tya}:
\begin{eqnarray}
\label{response}
V_2&\approx & -\kappa_2\varepsilon_2\cr
V_3&\approx & -\kappa_3\varepsilon_3,
\end{eqnarray}
where $\kappa_2$ and $\kappa_3$ are response coefficients~\cite{Teaney:2012ke}. 
These coefficients are real, by parity symmetry ($\phi\to -\phi$), and positive. 
The minus sign in Eq.~(\ref{response}), which is often included in the definition of $\varepsilon_n$, is due to the fact that the conversion from shape anisotropy to anisotropic flow involves a gradient. 
In other terms, it means that an almond elongated along the $y$ axis generates elliptic flow along the $x$ axis, as illustrated in Fig.~\ref{fig:b}. 
Equation~(\ref{response}) is in complex form~\cite{Gardim:2011xv}, which implies that the phase of $V_n$ coincides with that of $\varepsilon_n$, again to a good approximation. 

The response coefficients $\kappa_n$ depends on details of the hydrodynamic model (most notably, on transport coefficients of the quark-gluon plasma, and on how the freeze-out of the fluid into hadrons is modeled), and they are not known precisely. 
The important point for nuclear structure studies is that, for a given colliding energy $\sqrt{s_{NN}}$, they depend mildly on the nuclei involved in the collision. 
Let me explain why. 
I have mentioned at the end of Sec.~\ref{sec:intro} that the temperature of the quark-gluon plasma formed in the collision depends on $\sqrt{s_{NN}}$, but little on the colliding nuclei and on the collision centrality. 
Therefore, changing the colliding nuclei essentially amounts, in a first approximation, to a scale transformation $(x,y,z,t)\to(\lambda x,\lambda y,\lambda z,\lambda t)$, where $\lambda$ is a constant. 
The initial anisotropies $\varepsilon_n$ are invariant under this scale transformation. 
Ideal hydrodynamics is scale invariant, so that a solution of the equations is mapped onto another solution upon rescaling.  
Therefore, the response coefficients $\kappa_n$ would be independent of the colliding nuclei if the evolution was described by ideal hydrodynamics, and if a nucleus was exactly mapped onto another nucleus by a scale transformation. 
Scale invariance is broken by viscous corrections to ideal hydrodynamics. 
They suppress the response coefficients $\kappa_2$ and $\kappa_3$ by a fraction which is proportional to the shear and bulk viscosities, and inversely proportional to the system size (Reynolds number scaling) \cite{Gardim:2020mmy}.\footnote{The viscous suppression is roughly twice larger for $v_3$ than for $v_2$~\cite{Gardim:2022vys}.} 
This induces a dependence on system size, but a mild one. 
The same arguments apply to the variation of $\kappa_n$ on the collision centrality, which is mild. 

Therefore, the dependence of $v_2$ and $v_3$ on colliding nuclei and on collision centrality is mostly driven by that of the initial anisotropies $\varepsilon_2$ and $\varepsilon_3$. 
Note that higher harmonics,  $v_4$, $v_5$, $v_6$, $v_7$, are not described by simple linear response, because they are coupled to elliptic flow, which induces large nonlinear response terms~\cite{Gardim:2011xv,Yan:2015jma,ALICE:2017fcd,ALICE:2020sup,STAR:2020gcl}. 

\subsection{$\varepsilon_n$ is sensitive to nuclear structure}
\label{sec:structure2}

\begin{figure}[tb]
\begin{center}
\includegraphics[width=\linewidth]{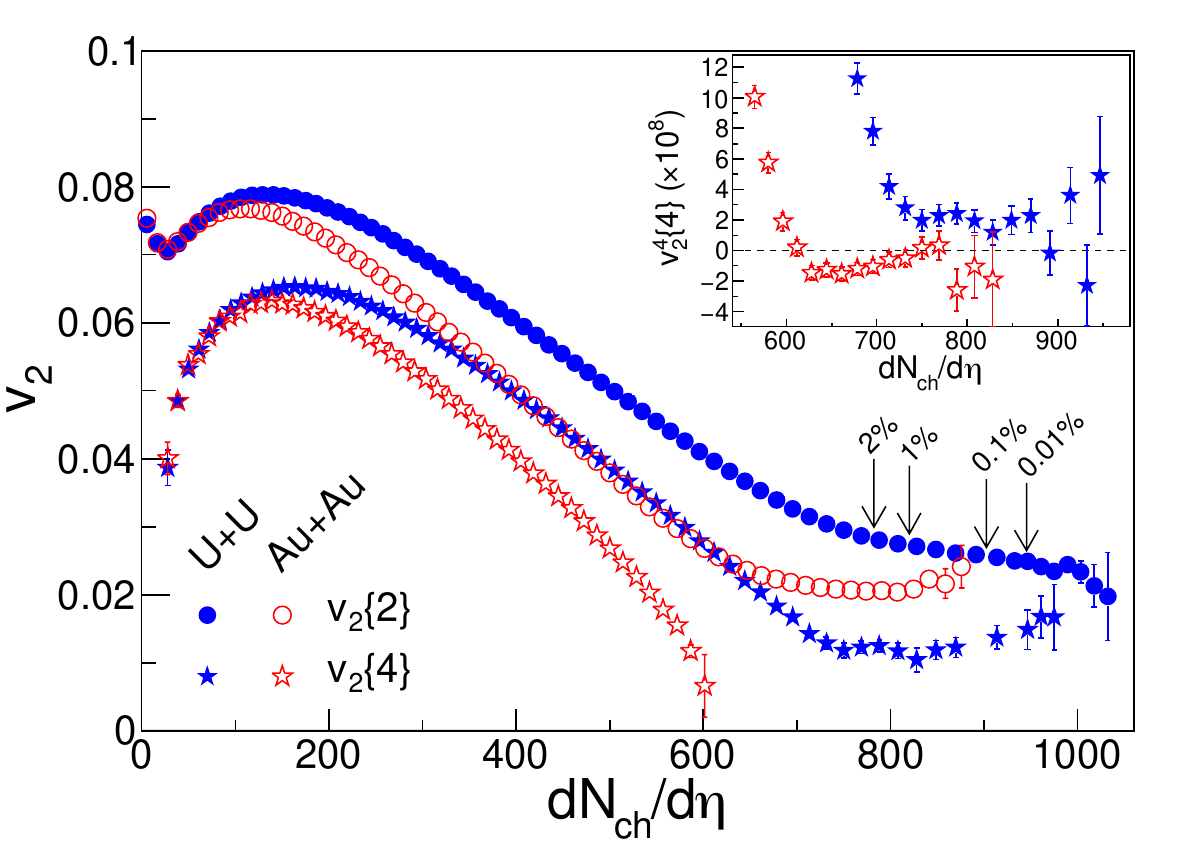}
\end{center}
\caption{Elliptic flow in $^{197}$Au+$^{197}$Au collisions at $\sqrt{s_{NN}}=200$~GeV (open symbols) and  $^{238}$U+$^{238}$U   collisions at $\sqrt{s_{NN}}=193$~GeV (full symbols). 
Circles: $v_2\{2\}$, defined by Eqs.~(\ref{defc22}) and (\ref{defv22}).
Stars: $v_2\{4\}$, defined by Eqs.~(\ref{cumul4}) and (\ref{defv24}). 
This figure is borrowed from Ref.~\cite{STAR:2015mki}.
 }
\label{fig:UU}
\end{figure}


Fig.~\ref{fig:UU} displays the dependence of $v_2\{2\}$ on event multiplicity in $^{197}$Au+$^{197}$Au (open circles) and  $^{238}$U+$^{238}$U collisions (filled circles).\footnote{ 
The Au+Au results are equivalent to those shown in Fig.~\ref{fig:STAR}, but they are of much better  quality, as many more data were available, and the analysis methods had also been refined. The larger values of $v_2$ are due to a combination of three factors: 1. The more recent analysis exludes particles with $0.1<p_t<0.2$~GeV$/c$. Particles in this range have $v_2\approx 0$, therefore, one increases the average $v_2$ by excluding them. 2.The older analysis uses the event-plane method, which yields results slightly smaller than $v_2\{2\}$ in the presence of flow fluctuations~\cite{Ollitrault:2009ie}. 3. The recent analysis is done at a slightly higher collision energy.  }
These data represent essentially, up to a multiplicative constant, the variation of $\varepsilon_2\{2\}$, which is largest for the lowest values of the charged multiplicity, corresponding to large values of $b$ where the almond in Fig.~\ref{fig:b} is elongated, and smallest in central collisions, where the ellipticity is only generated by fluctuations. 

One immediately sees that $v_2\{2\}$ is larger in U+U collisions than in Au+Au collisions, implying that $\varepsilon_2\{2\}$ is larger. 
This is a direct consequence of the large quadrupole deformation of the $^{238}$U nucleus in its ground state~\cite{Giacalone:2021udy}. 
Each of the two colliding nuclei has a random orientation with respect to the collision axis, and the quadrupole deformation is in general smaller once projected on the transverse plane. 
Still, on average, the deformation increases the ellipticity of the density profile.

\begin{figure}[ht]
    \centering
\includegraphics[width=\linewidth]{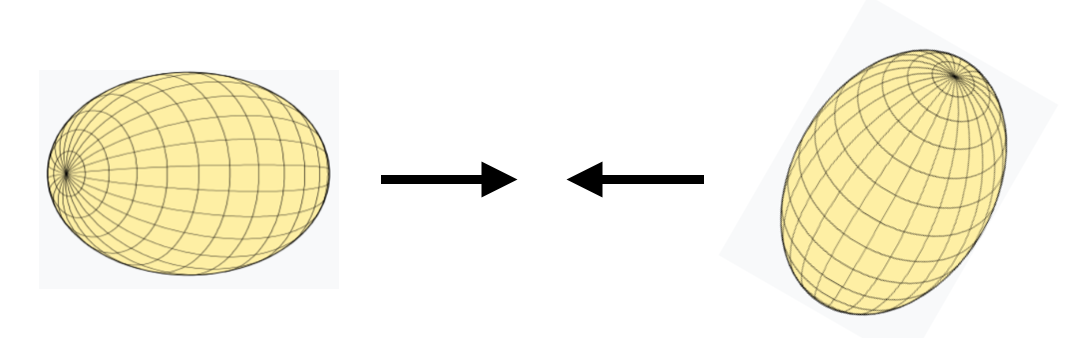}
\includegraphics[width=\linewidth]{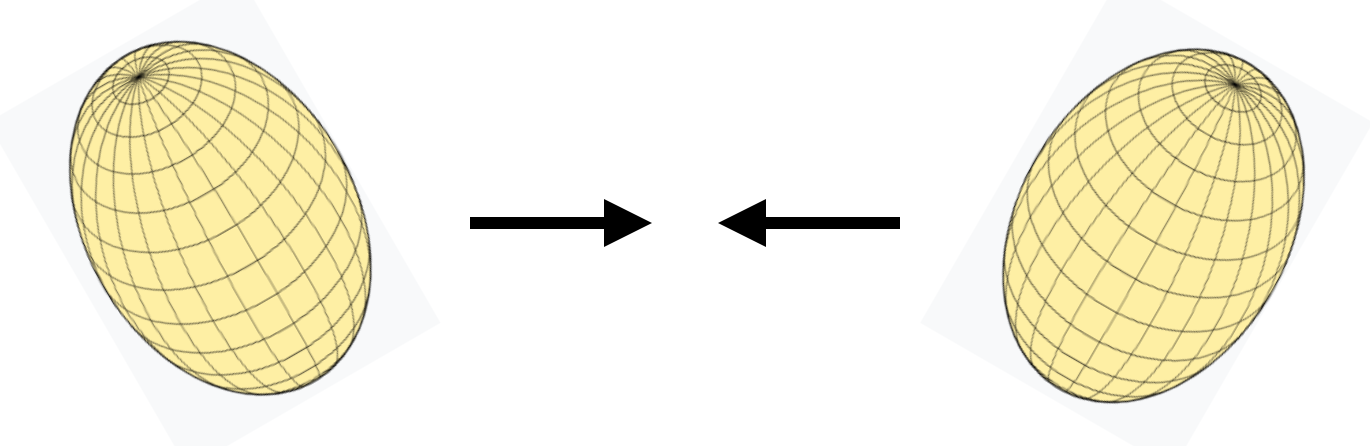}
    \caption{Schematic representation of collisions between two prolate nuclei at zero impact parameter, where Lorentz contraction has been omitted for clarity. In the first case (top), the nuclei do not fully overlap at the time of impact, which results in a smaller multiplicity than in the second case (bottom) where overlap is complete. 
    In high-multiplicity collisions between deformed nuclei, the first nucleus can have any orientation relative to the collision axis, but the second nucleus must be oriented in such a way that it fully overlaps.}
    \label{fig:orientation}
\end{figure}
The relative difference between Au+Au and U+U collisions is largest for high multiplicities. 
This can easily be understood. 
As explained in Sec.~\ref{sec:intro}, the multiplicity is proportional to the number of participant nucleons. 
It is largest if the impact parameter is close to 0 {\it and\/} if the two nuclei fully overlap, as depicted in Fig.~\ref{fig:orientation}. 
In such a situation, the collision zone has the same shape as a single nucleus, projected onto the transverse plane. 
Detailed comparison of U+U data with theoretical models~\cite{Moreland:2014oya,Schenke:2014tga} has proven that in this configuration, the multiplicity depends little on the angle $\theta$ between the symmetry axis of the nucleus and the collision axis, and is essentially the same if $\theta=0$ (tip-tip collision) as if $\theta=\pi/2$ (body-body collision).\footnote{Therefore, the events with the highest multiplicities can have any  value of $\theta$. 
The ellipticity $\varepsilon_2$ is minimal  for a tip-tip collision and maximal for a body-body collision, in which the deformation of the $^{238}$U nucleus shows up. 
The net effect, after averaging over $\theta$, is that this deformation increases $\varepsilon_2$.} 
The quadrupole deformation results in a larger ellipticity in U+U than in Au+Au collisions, where $\varepsilon_2$ is mostly due to local density fluctuations. 
Similarly, the observation of a larger $v_2$ in central $^{129}$Xe+$^{129}$Xe than in central $^{208}$Pb+$^{208}$Pb at the LHC~\cite{ALICE:2018lao,CMS:2019cyz,ATLAS:2019dct} provided the first direct experimental evidence that $^{129}$Xe has a significant quadrupole deformation.\footnote{One generally expects that $v_2$ is larger in central collisions if nuclei are smaller, due to larger fluctuations, as was observed in Cu+Cu collisions at RHIC (Sec.~\ref{sec:puzzles}). The statement  here is a quantitative one, that the observed increase of $v_2$ in Xe+Xe collisions relative to Pb+Pb is larger than one would expect just from the different nuclear size.}

Generally, collisions with the highest multiplicity are the best place to study effects of nuclear deformation, because the two nuclei essentially appear as one. 
An octupole deformation will typically lead to a triangular collision zone, thus increasing $\varepsilon_3$. 
$^{96}$Zr+$^{96}$Zr collisions at RHIC~\cite{STAR:2021mii} provided clear evidence that $^{96}$Zr has an octupole deformation, leading to a larger $v_3\{2\}$ in high-multiplicity collisions~\cite{Zhang:2021kxj} than in $^{96}$Ru+$^{96}$Ru (where $^{96}$Ru does not have an octupole deformation). 
This octupole deformation is also suggested by recent nuclear structure studies~\cite{Rong:2022qez}. 
Higher-order deformations can also be detected: 
Evidence that $^{238}$U possesses hexadecapole deformation~\cite{Ryssens:2023fkv} has recently been inferred from RHIC data.

Another quantity that can be constrained with heavy ions is the skin depth, which characterizes how fast the nuclear density decreases at the edge of the nucleus~\cite{Miller:2007ri}.  
In a non-central collision, such as depicted in Fig.~\ref{fig:b}, a smaller skin depth results in an almond with sharper edges, which increases the ellipticity $\varepsilon_2$~\cite{Xu:2021vpn,Jia:2022qgl}, hence a larger $v_2\{2\}$. 
The precision on the neutron skin of $^{208}$Pb from LHC data has recently been argued~\cite{Giacalone:2023cet} to be comparable with that of the dedicated PREX experiment~\cite{PREX:2021umo}. 

Relating quantitatively heavy-ion data to nuclear structure implies that one knows how the initial energy density $\epsilon(x,y)$, which defines initial anisotropies, is related to the thickness functions of colliding nuclei $T_A(x,y)$ and $T_B(x,y)$. 
The most common approach is phenomenological and involves rather general parametrizations, which are then adjusted to experimental data~\cite{Moreland:2014oya}. 
It is interesting to note that the parameters of the initial density are, among all the parameters of hydrodynamic calculations, those which are the most strongly constrained by global theory to data comparisons~\cite{JETSCAPE:2020mzn,Bernhard:2016tnd,Nijs:2020roc}. 
In recent years, a clear case has emerged that heavy-ion collisions are able to constrain the structure of colliding nuclei~\cite{Bally:2022vgo} better that they can constrain the properties of the quark-gluon plasma (typically its shear and bulk viscosity, which are still poorly known). 
However, this would still be considered a provocative statement in the heavy-ion community. 
It should be supported by global analyses, in which the deformations of the colliding nuclei are extracted from experimental data with a well-defined theoretical uncertainty.

\section{Probing anisotropy with cumulants}
\label{sec:cumulants}

\subsection{Higher-order correlations}

In Secs.~\ref{sec:v2} and \ref{sec:ridge}, we have introduced measures of azimuthal anisotropy based on pair correlations. 
Now, the particle multiplicity in heavy-ion collisions is so large that one can easily correlate more than two particles from the same event. 
Higher-order correlations are golden signatures of collective effects and have been much studied, both theoretically and experimentally, over the last two decades. 

They were first applied~\cite{STAR:2002hbo} to the measurement of elliptic flow in non-central collisions. 
The idea was to consider, rather than one pair, two pairs in the same event, with azimuthal angles $(\phi_1,\phi_2)$ for one pair, and $(\phi_3,\phi_4)$ for the other pair.  
Instead of averaging $e^{2i(\phi_1-\phi_2)}$ over all pairs, as in Eq.~(\ref{defc22}), one multiplies the contribution of each pair before taking the average, that is, one averages $e^{2i(\phi_1-\phi_2)}e^{2i(\phi_3-\phi_4)}=e^{2i(\phi_1-\phi_2+\phi_3-\phi_4)}$ over all $4$-plets. 
One then applies the flow paradigm, that particles in a single event are emitted independently, with an anisotropic distribution $P(\phi)$ given by Eq.~(\ref{fourier}). 
For a single particle, the expectation value of $e^{2i\phi}$ is $V_2$, according to Eq.~(\ref{inversefourier}), and the expectation value of $e^{-2i\phi}$ is $V_{-2}=V_2^*$. 
Therefore, the expectation value of $e^{2i(\phi_1-\phi_2+\phi_3-\phi_4)}$ in a single event is $\left| V_2\right|^4$. 
Averaging over a large number of events at fixed centrality, one finally obtains
\begin{equation}
\label{fourcorrelation}
\left\langle e^{2i(\phi_1-\phi_2+\phi_3-\phi_4)} \right\rangle=\left\langle \left| V_2\right|^4 \right\rangle . 
\end{equation}
If $V_2$ was identical for all events in a centrality class, as in the old flow picture of Sec.~\ref{sec:v2}, the right-hand side would be $(v_2)^4$, and information from the four-particle correlation in Eq.~(\ref{fourcorrelation}) would be redundant with that from the pair correlation in Eq.~(\ref{factorization}). 
But the modern flow picture of Sec~\ref{sec:ridge} has taught us that $V_2$ fluctuates event to event. 
Then, higher-order correlators give access to higher-order moments of the distribution of $\left|V_2\right|^2$~\cite{Bhalerao:2014xra}, from which one can reconstruct the full probability distribution of $\left|V_2\right|$~\cite{Mehrabpour:2018kjs}.

\subsection{Cumulants}
In practice, however, experiments do not measure directly moments such as (\ref{fourcorrelation}), but combination of moments known as cumulants. 
The reason is mostly historical. 
Higher-order correlations were introduced originally in order to separate flow from nonflow correlations, and cumulants precisely achieve this separation~\cite{Borghini:2000sa}. 
It was realized only a few years later~\cite{PHENIX:2003qra} that there is a simpler method for suppressing nonflow, which  is to implement a rapidity gap in pair correlations, as explained in Sec.~\ref{sec:vn2}.  

I will illustrate cumulants using the best-known example, which is the four-particle cumulant. 
Let us first assume that particles are produced in bunches of correlated pairs. 
Pairwise correlations induce a four-particle correlation, which is obtained by pairing the four particles in all possible ways: $(1,2)(3,4)$ or  $(1,3)(2,4)$ or $(1,4)(2,3)$.  
Each pairing gives a different contribution to the four-particle correlator (\ref{fourcorrelation}), which is decomposed as: 
\begin{eqnarray}
\label{decomp1}
\left\langle e^{2i(\phi_1-\phi_2+\phi_3-\phi_4)}\right\rangle&=&\left\langle e^{2i(\phi_1-\phi_2)}\right\rangle\left\langle e^{2i(\phi_3-\phi_4)}\right\rangle\cr
&&+\left\langle e^{2i(\phi_1+\phi_3)}\right\rangle\left\langle e^{2i(-\phi_2-\phi_4)}\right\rangle\cr
&&+\left\langle e^{2i(\phi_1-\phi_4)}\right\rangle\left\langle e^{2i(\phi_2-\phi_3)}\right\rangle.
\end{eqnarray}
Next, note that the second term in the right-hand side vanishes due to azimuthal symmetry. 
What I mean is that an event rotated by an angle $\alpha$ around the collision axis (so that $\phi_i$ is replaced with $\phi_i+\alpha$) is equally probable.
The first and third term only involve relative angles and are independent of  $\alpha$, while the second term vanishes upon averaging over $\alpha$. 
Thus the decomposition reduces to: 
\begin{eqnarray}
\label{decomp2}
\left\langle e^{2i(\phi_1-\phi_2+\phi_3-\phi_4)}\right\rangle&=&\left\langle e^{2i(\phi_1-\phi_2)}\right\rangle\left\langle e^{2i(\phi_3-\phi_4)}\right\rangle\cr
&&+\left\langle e^{2i(\phi_1-\phi_4)}\right\rangle\left\langle e^{2i(\phi_2-\phi_3)}\right\rangle.
\end{eqnarray}
Now, the {\it cumulant\/} $c_2\{4\}$ is defined~\cite{Borghini:2001vi} by subtracting the right-hand side from the left-hand side:
\begin{eqnarray}
\label{cumul4}
c_2\{4\}&\equiv& \left\langle e^{2i(\phi_1-\phi_2+\phi_3-\phi_4)}\right\rangle\cr 
&&-\left\langle e^{2i(\phi_1-\phi_2)}\right\rangle\left\langle e^{2i(\phi_3-\phi_4)}\right\rangle\cr
&&-\left\langle e^{2i(\phi_1-\phi_4)}\right\rangle\left\langle e^{2i(\phi_2-\phi_3)}\right\rangle.
\end{eqnarray}
This cumulant, which is routinely measured in heavy-ion collisions, corresponds to the {\it genuine\/} four-particle correlation, which remains after contributions from underlying pair correlations have been subtracted. 

Instead of assuming that particles are correlated in pairs, let us now assume that the flow paradigm applies. 
Then, the first term in Eq.~(\ref{cumul4}) is given by Eq.~(\ref{fourcorrelation}), while the pair correlation is given by Eq.~(\ref{defcn2}), and one obtains: 
\begin{equation}
\label{cumul4flow}
c_2\{4\}=\left\langle \left| V_2\right|^4 \right\rangle-2 \left\langle \left| V_2\right|^2 \right\rangle^2.
\end{equation}
If $V_2$ was identical in all events, as in the old flow paradigm of Sec.~\ref{sec:v2}, this would reduce to 
\begin{equation}
\label{cumul4flow2}
c_2\{4\}=-(v_2)^4.
\end{equation}
Thus a constant $v_2$ results in a negative $c_2\{4\}$.  

One denotes by $v_2\{4\}$ the estimate of $v_2$ that one would obtain from Eq.~(\ref{cumul4flow2})~\cite{Borghini:2001vi}: 
\begin{equation}
\label{defv24}
v_2\{4\}\equiv \left(-c_2\{4\}\right)^{1/4}. 
\end{equation}
Since $V_2$ fluctuates event-to-event, $v_2\{4\}$ measures a non-trivial combination of moments, as seen by inserting Eq.~(\ref{cumul4flow}) into Eq.~(\ref{defv24}). 

If nonflow effects have been correctly removed and Eqs.~(\ref{defvn2}) and (\ref{cumul4flow}) both apply, it follows that $v_2\{4\}<v_2\{2\}$, as a consequence of the inequality $\left\langle\left| V_2\right|^4\right\rangle>\left\langle\left| V_2\right|^2\right\rangle^2$. 
This ordering is clearly seen in Fig.~\ref{fig:UU}. 

Note also that if fluctuations are large enough that $\left\langle\left| V_2\right|^4\right\rangle>2\left\langle\left| V_2\right|^2\right\rangle^2$, then $c_2\{4\}>0$ and $v_2\{4\}$ is undefined. 
This is observed in Au+Au collisions at large multiplicity (see inlay in Fig.~\ref{fig:UU}), and also in central Pb+Pb collisions at the LHC~\cite{ATLAS:2019peb}. 

The relative difference between $v_2\{4\}$ and $v_2\{2\}$ is small when the fluctuations of $|V_2|$ are small.  
One sees in Fig.~\ref{fig:UU} that it the case for moderate values of the multiplicity, corresponding to non-central collisions where the static deformation discussed in Sec.~\ref{sec:v2} is the dominant source of elliptic flow.  
If nuclei are not strongly deformed, it turns out that $v_2\{4\}$ coincides, to a very good approximation~\cite{Bhalerao:2006tp}, with the original definition of $v_2$ as the average $\cos 2\phi$ relative to the direction of impact parameter, used in Sec.~\ref{sec:v2}. 
The reason is that the subtraction in Eq.~(\ref{cumul4}) removes not only nonflow correlations, but also Gaussian flow fluctuations~\cite{Voloshin:2007pc}, and flow fluctuations are Gaussian to a good approximation. 
Even though cumulants were introduced before flow fluctuations were discovered, they end up being useful also to study flow fluctuations.

Cumulants can be defined for other Fourier harmonics, simply by replacing $V_2$ with $V_n$ in the above discussion. 
The ALICE Collaboration measured $v_3\{4\}$ very early on~\cite{ALICE:2011ab}, and the ATLAS Collaboration measured $c_4\{4\}$  a few years later~\cite{ATLAS:2014qxy}.  
$c_4\{4\}$ changes sign changes sign for large impact parameters~\cite{ATLAS:2019peb}, so that $v_4\{4\}$ cannot always be defined. 
This change of sign is induced by the nonlinear response term mentioned at the end of  Sec.~\ref{sec:epsn}~\cite{Giacalone:2016mdr}.

The discussion can also be generalized to higher-order cumulants, which have been precisely measured up to order 10 for $v_2$~\cite{CMS:2022umz}.
 $v_2\{6\}$ and $v_2\{8\}$ are found to be almost equal to $v_2\{4\}$ (they all measure the static elliptic deformation), and the tiny relative differences (at the percent level between $v_2\{4\}$ and $v_2\{6\}$, and at the per mille level between $v_2\{6\}$ and $v_2\{8\}$) are used to probe the small non-Gaussianities (skewness~\cite{Giacalone:2016eyu,CMS:2017glf,ALICE:2018rtz} and kurtosis~\cite{CMS:2022umz,Bhalerao:2018anl}) of elliptic flow fluctuations, and compare them with predictions from hydrodynamic models. 
 
Mixed cumulants of various types, involving several harmonics, have also been measured in fixed target experiments \cite{NA49:2003njx}, at RHIC~\cite{STAR:2003xyj} and at the LHC~\cite{ATLAS:2014ndd,ALICE:2016kpq}. 
Finally, a new type of cumulant, measuring the correlation between the azimuthal anisotropy and the average transverse momentum, has been introduced by Bo\.zek~\cite{Bozek:2016yoj} and much studied recently~\cite{ATLAS:2019pvn,ALICE:2021gxt,ATLAS:2022dov}

\subsection{Interest for nuclear structure} 

We have seen in Sec.~\ref{sec:structure2} that central collisions between identical deformed nuclei correspond to configurations where the two nuclei fully overlap, but can have various orientations with respect to the collision axis. 
Different orientations typically yield different $V_n$. 
For example, for a nucleus with large quadrupole deformation such as $^{238}$U,    $\left|V_2\right|$ is larger in a body-body collision, where the overlap area has the same deformation as the nucleus itself, than in a tip-tip collision where the overlap area is circular. 
In this case, fluctuations of $\left|V_2\right|$ mostly stem from fluctuations in the orientation. 
Due to the deformation, $c_2\{4\}$ remains negative even in the most central collisions~\cite{Giacalone:2018apa}, so that $v_2\{4\}$ can always be defined (see inlay in Fig.~\ref{fig:UU}). 

In general, cumulants are useful for studying nuclear deformation because the various orientations of nuclei during the collision result in fluctuations of the azimuthal anistropy~\cite{Jia:2021tzt,Jia:2021qyu}. 
I will only mention, as an illustration, that Bo\.zek's mixed cumulant was instrumental to demonstrate the impact of the triaxial structure of $^{129}$Xe on high-energy data~\cite{Bally:2021qys}, and has recently been suggested as a probe of the triaxiality of $^{197}$Au~\cite{Bally:2023dxi}. 


\section{Conclusions}
\label{sec:conclusion}

In summary, azimuthal anisotropies provide a detailed two-dimensional image of the collision zone in a nucleus-nucleus collision at very high energies. 
This is fundamentally a dynamic picture, since the collision occurs on time scale so short that it captures each of the nuclei in a specific quantum state. 
Relating this image of the collision zone to that of a single nucleus is not straightforward for two reasons. 
First, the collision involves two nuclei. 
Second, the nucleus is a three-dimensional object, but the Lorentz contraction projects it onto a two-dimen\-sional image. 
The second difficulty is purely technical, not conceptual. 
The first difficulty is the most serious one, and there is at present no solid understanding of how the densities of the two nuclei combine to produce quark-gluon matter in the collision zone. 
Despite this theoretical uncertainty, global theory-to-data comparisons are already able to strongly constrain the correspondence between incoming nuclear density and pro\-duced matter density. 
Nuclear structure studies from heavy-ion collisions are still in their infancy, but a wealth of new results are within reach, provided that accelerators deliver a wide variety of nuclear beams at sufficiently high energy.


\begin{thebibliography}{99}
\bibitem{Busza:2018rrf}
W.~Busza, K.~Rajagopal, W.~van der Schee,
Ann. Rev. Nucl. Part. Sci. \textbf{68} (2018), 339-376
https://arxiv.org/abs/1802.04801 [hep-ph].

\bibitem{Hadjidakis:2018ifr}
C.~Hadjidakis, D.~Kiko\l{}a, J.~P.~Lansberg, L.~Massacrier, M.~G.~Echevarria, A.~Kusina, I.~Schienbein, J.~Seixas, H.~S.~Shao and A.~Signori, \textit{et al.}
Phys. Rept. \textbf{911} (2021), 1-83
https://arxiv.org/abs/1807.00603 [hep-ex].

\bibitem{Bjorken:1968dy}
J.~D.~Bjorken,
Phys. Rev. \textbf{179} (1969), 1547-1553

\bibitem{Kajantie:1987pd}
K.~Kajantie, P.~V.~Landshoff and J.~Lindfors,
Phys. Rev. Lett. \textbf{59} (1987), 2527

\bibitem{Krasnitz:1998ns}
A.~Krasnitz and R.~Venugopalan,
Nucl. Phys. B \textbf{557} (1999), 237
https://arxiv.org/abs/hep-ph/9809433 [hep-ph].

\bibitem{Eskola:1999fc}
K.~J.~Eskola, K.~Kajantie, P.~V.~Ruuskanen and K.~Tuominen,
Nucl. Phys. B \textbf{570} (2000), 379-389 \\
https://arxiv.org/abs/hep-ph/9909456 [hep-ph].

\bibitem{Eskola:2001bf}
K.~J.~Eskola, P.~V.~Ruuskanen, S.~S.~Rasanen and K.~Tuominen,
Nucl. Phys. A \textbf{696} (2001), 715-728
https://arxiv.org/abs/hep-ph/0104010 [hep-ph].

\bibitem{Schenke:2012wb}
B.~Schenke, P.~Tribedy and R.~Venugopalan,
Phys. Rev. Lett. \textbf{108} (2012), 252301
https://arxiv.org/abs/1202.6646 [nucl-th].

\bibitem{Moreland:2014oya}
J.~S.~Moreland, J.~E.~Bernhard and S.~A.~Bass,
Phys. Rev. C \textbf{92} (2015) no.1, 011901
https://arxiv.org/abs/1412.4708 [nucl-th].

\bibitem{Nagle:2018ybc}
J.~L.~Nagle and W.~A.~Zajc,
Phys. Rev. C \textbf{99} (2019) no.5, 054908
https://arxiv.org/abs/1808.01276 [nucl-th].

\bibitem{Schenke:2020mbo}
B.~Schenke, C.~Shen and P.~Tribedy,
Phys. Rev. C \textbf{102} (2020) no.4, 044905
https://arxiv.org/abs/2005.14682 [nucl-th].

\bibitem{Giacalone:2023hwk}
G.~Giacalone,
https://arxiv.org/abs/2305.19843    

\bibitem{Braun-Munzinger:2003pwq}
P.~Braun-Munzinger, K.~Redlich and J.~Stachel,
``Particle production in heavy ion collisions,'' in Quark Gluon Plasma 3, eds. R. C. Hwa and Xin-Nian Wang, World Scientific Publishing
https://arxiv.org/abs/nucl-th/0304013 [nucl-th].

\bibitem{ALICE:2016fbt}
J.~Adam \textit{et al.} [ALICE],
Phys. Lett. B \textbf{772} (2017), 567-577
https://arxiv.org/abs/1612.08966 [nucl-ex].

\bibitem{Das:2017ned}
S.~J.~Das, G.~Giacalone, P.~A.~Monard and J.~Y.~Ollitrault,
Phys. Rev. C \textbf{97} (2018) no.1, 014905
https://arxiv.org/abs/1708.00081 [nucl-th].

\bibitem{Miller:2007ri}
M.~L.~Miller, K.~Reygers, S.~J.~Sanders and P.~Steinberg,
Ann. Rev. Nucl. Part. Sci. \textbf{57} (2007), 205-243
https://arxiv.org/abs/nucl-ex/0701025 [nucl-ex].

\bibitem{Eremin:2003qn}
S.~Eremin and S.~Voloshin,
Phys. Rev. C \textbf{67} (2003), 064905
https://arxiv.org/abs/nucl-th/0302071 [nucl-th].

\bibitem{Bialas:2006kw}
A.~Bialas and A.~Bzdak,
Phys. Lett. B \textbf{649} (2007), 263-268
[erratum: Phys. Lett. B \textbf{773} (2017), 681-681]
https://arxiv.org/abs/nucl-th/0611021 [nucl-th].

\bibitem{PHENIX:2013ehw}
S.~S.~Adler \textit{et al.} [PHENIX],
Phys. Rev. C \textbf{89} (2014) no.4, 044905
https://arxiv.org/abs/1312.6676 [nucl-ex].

\bibitem{Loizides:2016djv}
C.~Loizides,
Phys. Rev. C \textbf{94} (2016) no.2, 024914
https://arxiv.org/abs/1603.07375 [nucl-ex].

\bibitem{ALICE:2013hur}
B.~Abelev \textit{et al.} [ALICE],
Phys. Rev. C \textbf{88} (2013) no.4, 044909
https://arxiv.org/abs/1301.4361 [nucl-ex].

\bibitem{Yousefnia:2021cup}
K.~V.~Yousefnia, A.~Kotibhaskar, R.~Bhalerao and J.~Y.~Ollitrault,
Phys. Rev. C \textbf{105} (2022) no.1, 014907
https://arxiv.org/abs/2108.03471 [nucl-th].

\bibitem{Poskanzer:1998yz}
A.~M.~Poskanzer and S.~A.~Voloshin,
Phys. Rev. C \textbf{58} (1998), 1671-1678
https://arxiv.org/abs/nucl-ex/9805001 [nucl-ex].

\bibitem{Giacalone:2019pca}
G.~Giacalone,
Phys. Rev. Lett. \textbf{124} (2020) no.20, 202301
https://arxiv.org/abs/1910.04673 [nucl-th].

\bibitem{Gardim:2019xjs}
F.~G.~Gardim, G.~Giacalone, M.~Luzum and J.~Y.~Ollitrault,
Nature Phys. \textbf{16} (2020) no.6, 615-619
https://arxiv.org/abs/1908.09728 [nucl-th].

\bibitem{ALICE:2018hza}
S.~Acharya \textit{et al.} [ALICE],
Phys. Lett. B \textbf{788} (2019), 166-179
https://arxiv.org/abs/1805.04399 [nucl-ex].

\bibitem{PHOBOS:2000wxz}
B.~B.~Back \textit{et al.} [PHOBOS],
Phys. Rev. Lett. \textbf{85} (2000), 3100-3104
https://arxiv.org/abs/hep-ex/0007036 [hep-ex].

\bibitem{STAR:2000ekf}
K.~H.~Ackermann \textit{et al.} [STAR],
Phys. Rev. Lett. \textbf{86} (2001), 402-407
https://arxiv.org/abs/nucl-ex/0009011 [nucl-ex].

\bibitem{Ollitrault:1992bk}
J.~Y.~Ollitrault,
Phys. Rev. D \textbf{46} (1992), 229-245

\bibitem{E877:1994plr}
J.~Barrette \textit{et al.} [E877],
Phys. Rev. Lett. \textbf{73} (1994), 2532-2535
https://arxiv.org/abs/hep-ex/9405003 [hep-ex].

\bibitem{NA49:1997qey}
H.~Appelshauser \textit{et al.} [NA49],
Phys. Rev. Lett. \textbf{80} (1998), 4136-4140
https://arxiv.org/abs/nucl-ex/9711001 [nucl-ex].

\bibitem{Ollitrault:1993iw}
J.~Y.~Ollitrault,
``Anisotropy: A new signature of transverse collective flow?,''
in Proceedings of 2nd International Conference on Physics and Astrophysics of Quark Gluon Plasma (ICPAQGP 1993), Calcutta, January 19-23, 1993, editors, Bikash Sinha, Yogendra Pathak Viyogi, Sibaji Raha, World Scientific, 1993. 

\bibitem{Kolb:2000sd}
P.~F.~Kolb, J.~Sollfrank and U.~W.~Heinz,
Phys. Rev. C \textbf{62} (2000), 054909
https://arxiv.org/abs/hep-ph/0006129 [hep-ph].

\bibitem{Romatschke:2007mq}
P.~Romatschke and U.~Romatschke,
Phys. Rev. Lett. \textbf{99} (2007), 172301
https://arxiv.org/abs/0706.1522 [nucl-th].

\bibitem{Borsanyi:2013bia}
S.~Borsanyi, Z.~Fodor, C.~Hoelbling, S.~D.~Katz, S.~Krieg and K.~K.~Szabo,
Phys. Lett. B \textbf{730} (2014), 99-104
https://arxiv.org/abs/1309.5258 [hep-lat].

\bibitem{JETSCAPE:2020mzn}
D.~Everett \textit{et al.} [JETSCAPE],
Phys. Rev. C \textbf{103} (2021) no.5, 054904
https://arxiv.org/abs/2011.01430 [hep-ph].

\bibitem{Voloshin:1994mz}
S.~Voloshin and Y.~Zhang,
Z. Phys. C \textbf{70} (1996), 665-672
https://arxiv.org/abs/hep-ph/9407282 [hep-ph].

\bibitem{Jacak:2010zz}
B.~Jacak and P.~Steinberg,
Phys. Today \textbf{63N5} (2010), 39-43

\bibitem{Gardim:2022vys}
F.~G.~Gardim and J.~Y.~Ollitrault,
Acta Phys. Polon. Supp. \textbf{16} (2023) no.1, 88
https://arxiv.org/abs/2207.08692 [nucl-th].

\bibitem{Velkovska}
J. Velkovska, S. V. Greene, 
Observing droplets of near-perfect fluid in high energy nuclear collisions (2021) 
https://www.innovationnewsnetwork.com/observing-droplets-of-near-perfect-fluid-in-high-energy-nuclear-collisions/12067/

\bibitem{Wang:1991qh}
S.~Wang, Y.~Z.~Jiang, Y.~M.~Liu, D.~Keane, D.~Beavis, S.~Y.~Chu, S.~Y.~Fung, M.~Vient, C.~Hartnack and H.~Stoecker,
Phys. Rev. C \textbf{44} (1991), 1091-1095

\bibitem{Borghini:2001vi}
N.~Borghini, P.~M.~Dinh and J.~Y.~Ollitrault,
Phys. Rev. C \textbf{64} (2001), 054901
https://arxiv.org/abs/nucl-th/0105040 [nucl-th].

\bibitem{Bilandzic:2010jr}
A.~Bilandzic, R.~Snellings and S.~Voloshin,
Phys. Rev. C \textbf{83} (2011), 044913
https://arxiv.org/abs/1010.0233 [nucl-ex].

\bibitem{STAR:2003xyj}
J.~Adams \textit{et al.} [STAR],
Phys. Rev. Lett. \textbf{92} (2004), 062301
[erratum: Phys. Rev. Lett. \textbf{127} (2021) no.6, 069901]
https://arxiv.org/abs/nucl-ex/0310029 [nucl-ex].


\bibitem{Ollitrault:1997vz}
J.~Y.~Ollitrault,
Nucl. Phys. A \textbf{638} (1998), 195-206
https://arxiv.org/abs/nucl-ex/9802005 [nucl-ex].

\bibitem{Ollitrault:1993ba}
J.~Y.~Ollitrault,
Phys. Rev. D \textbf{48} (1993), 1132-1139
https://arxiv.org/abs/hep-ph/9303247 [hep-ph].

\bibitem{Gutbrod:1989gh}
H.~H.~Gutbrod, K.~H.~Kampert, B.~Kolb, A.~M.~Poskanzer, H.~G.~Ritter, R.~Schicker and H.~R.~Schmidt,
Phys. Rev. C \textbf{42} (1990), 640-651


\bibitem{Brill:1993xh}
D.~Brill, C.~Bormann, Y.~Shin, J.~Stein, K.~Stiebing, R.~Stock, H.~Strobele, W.~Ahner, R.~Barth and M.~Cieslak, \textit{et al.}
Phys. Rev. Lett. \textbf{71} (1993), 336-339

\bibitem{E877:1997zjw}
J.~Barrette \textit{et al.} [E877],
Phys. Rev. C \textbf{56} (1997), 3254-3264
https://arxiv.org/abs/nucl-ex/9707002 [nucl-ex].

\bibitem{E895:1999ldn}
C.~Pinkenburg \textit{et al.} [E895],
Phys. Rev. Lett. \textbf{83} (1999), 1295-1298
https://arxiv.org/abs/nucl-ex/9903010 [nucl-ex].

\bibitem{Ollitrault:1997di}
J.~Y.~Ollitrault,
https://arxiv.org/abs/nucl-ex/9711003 [nucl-ex].

\bibitem{Danielewicz:1985hn}
P.~Danielewicz and G.~Odyniec,
Phys. Lett. B \textbf{157} (1985), 146-150
https://arxiv.org/abs/2109.05308 [nucl-th].

\bibitem{Ollitrault:2009ie}
J.~Y.~Ollitrault, A.~M.~Poskanzer and S.~A.~Voloshin,
Phys. Rev. C \textbf{80} (2009), 014904
https://arxiv.org/abs/0904.2315 [nucl-ex].

\bibitem{Luzum:2012da}
M.~Luzum and J.~Y.~Ollitrault,
Phys. Rev. C \textbf{87} (2013) no.4, 044907
https://arxiv.org/abs/1209.2323 [nucl-ex].

\bibitem{Aguiar:2001ac}
C.~E.~Aguiar, Y.~Hama, T.~Kodama and T.~Osada,
Nucl. Phys. A \textbf{698} (2002), 639-642
https://arxiv.org/abs/hep-ph/0106266 [hep-ph].

\bibitem{alverthesis}
Burak Han Alver, 
``Measurement of Non-flow Correlations and Elliptic Flow Fluctuations in Au+Au collisions at RHIC'', 
Massachusetts Institute of Technology 2010,
http://web.mit.edu/mithig/theses/Burak-Alver-thesis.pdf

\bibitem{PHOBOS:2005gex}
S.~Manly \textit{et al.} [PHOBOS],
Nucl. Phys. A \textbf{774} (2006), 523-526
https://arxiv.org/abs/nucl-ex/0510031 [nucl-ex].

\bibitem{PHOBOS:2006dbo}
B.~Alver \textit{et al.} [PHOBOS],
Phys. Rev. Lett. \textbf{98} (2007), 242302
https://arxiv.org/abs/nucl-ex/0610037 [nucl-ex].

\bibitem{STAR:2004wfp}
J.~Adams \textit{et al.} [STAR],
Phys. Rev. C \textbf{73} (2006), 064907
https://arxiv.org/abs/nucl-ex/0411003 [nucl-ex].

\bibitem{CMS:2012xss}
S.~Chatrchyan \textit{et al.} [CMS],
Eur. Phys. J. C \textbf{72} (2012), 2012
https://arxiv.org/abs/1201.3158 [nucl-ex].

\bibitem{Anderson:1972pca}
P.~W.~Anderson,
Science \textbf{177} (1972) no.4047, 393-396

\bibitem{Putschke:2007mi}
J.~Putschke [STAR],
J. Phys. G \textbf{34} (2007), S679-684
https://arxiv.org/abs/nucl-ex/0701074 [nucl-ex].

\bibitem{Putschkeridge}
J.~Putschke, private communication. 

\bibitem{Alver:2010gr}
B.~Alver and G.~Roland,
Phys. Rev. C \textbf{81} (2010), 054905
[erratum: Phys. Rev. C \textbf{82} (2010), 039903]
https://arxiv.org/abs/1003.0194 [nucl-th].

\bibitem{Luzum:2011mm}
M.~Luzum,
J. Phys. G \textbf{38} (2011), 124026
https://arxiv.org/abs/1107.0592 [nucl-th].

\bibitem{Giacalone:2016mdr}
G.~Giacalone, L.~Yan, J.~Noronha-Hostler and J.~Y.~Ollitrault,
J. Phys. Conf. Ser. \textbf{779} (2017) no.1, 012064
https://arxiv.org/abs/1608.06022 [nucl-th].

\bibitem{Giacalone:2022hnz}
G.~Giacalone,
https://arxiv.org/abs/2208.06839 [nucl-th].

\bibitem{Nijs:2023yab}
G.~Nijs and W.~van der Schee,
https://arxiv.org/abs/2304.06191 [nucl-th].

\bibitem{Bjorken:1982qr}
J.~D.~Bjorken,
Phys. Rev. D \textbf{27} (1983), 140-151

\bibitem{Acharya:2018lmh}
S.~Acharya \textit{et al.} [ALICE],
JHEP \textbf{07} (2018), 103
https://arxiv.org/abs/1804.02944 [nucl-ex].

\bibitem{Aaboud:2018ves}
M.~Aaboud \textit{et al.} [ATLAS],
Eur. Phys. J. C \textbf{78} (2018) no.12, 997
https://arxiv.org/abs/1808.03951 [nucl-ex].

\bibitem{Luzum:2012wu}
M.~Luzum and J.~Y.~Ollitrault,
Nucl. Phys. A \textbf{904-905} (2013), 377c-380c
https://arxiv.org/abs/1210.6010 [nucl-th].

\bibitem{Giannini:2022bkn}
A.~V.~Giannini \textit{et al.} [ExTrEMe],
Phys. Rev. C \textbf{107} (2023) no.4, 044907
https://arxiv.org/abs/2203.17011 [nucl-th].

\bibitem{Carzon:2020xwp}
P.~Carzon, S.~Rao, M.~Luzum, M.~Sievert and J.~Noronha-Hostler,
Phys. Rev. C \textbf{102} (2020) no.5, 054905
https://arxiv.org/abs/2007.00780 [nucl-th].

\bibitem{Teaney:2010vd}
D.~Teaney and L.~Yan,
Phys. Rev. C \textbf{83} (2011), 064904
https://arxiv.org/abs/1010.1876 [nucl-th].

\bibitem{Borghini:2000cm}
N.~Borghini, P.~M.~Dinh and J.~Y.~Ollitrault,
Phys. Rev. C \textbf{62} (2000), 034902
https://arxiv.org/abs/nucl-th/0004026 [nucl-th].

\bibitem{Retinskaya:2012ky}
E.~Retinskaya, M.~Luzum and J.~Y.~Ollitrault,
Phys. Rev. Lett. \textbf{108} (2012), 252302
https://arxiv.org/abs/1203.0931 [nucl-th].

\bibitem{ATLAS:2012at}
G.~Aad \textit{et al.} [ATLAS],
Phys. Rev. C \textbf{86} (2012), 014907
https://arxiv.org/abs/1203.3087 [hep-ex].

\bibitem{Miller:2003kd}
M.~Miller and R.~Snellings,
https://arxiv.org/abs/nucl-ex/0312008 [nucl-ex].

\bibitem{Majumder:2006wi}
A.~Majumder, B.~Muller and S.~A.~Bass,
Phys. Rev. Lett. \textbf{99} (2007), 042301
https://arxiv.org/abs/hep-ph/0611135 [hep-ph].

\bibitem{Dumitru:2007rp}
A.~Dumitru, Y.~Nara, B.~Schenke and M.~Strickland,
Phys. Rev. C \textbf{78} (2008), 024909
https://arxiv.org/abs/0710.1223 [hep-ph].

\bibitem{Casalderrey-Solana:2004fdk}
J.~Casalderrey-Solana, E.~V.~Shuryak and D.~Teaney,
J. Phys. Conf. Ser. \textbf{27} (2005), 22-31
https://arxiv.org/abs/hep-ph/0411315 [hep-ph].

\bibitem{Satarov:2005mv}
L.~M.~Satarov, H.~Stoecker and I.~N.~Mishustin,
Phys. Lett. B \textbf{627} (2005), 64-70
https://arxiv.org/abs/hep-ph/0505245 [hep-ph].

\bibitem{Ruppert:2005uz}
J.~Ruppert and B.~Muller,
Phys. Lett. B \textbf{618} (2005), 123-130
https://arxiv.org/abs/hep-ph/0503158 [hep-ph].

\bibitem{Koch:2005sx}
V.~Koch, A.~Majumder and X.~N.~Wang,
Phys. Rev. Lett. \textbf{96} (2006), 172302
https://arxiv.org/abs/nucl-th/0507063 [nucl-th].

\bibitem{Shuryak:2007fu}
E.~V.~Shuryak,
Phys. Rev. C \textbf{76} (2007), 047901
https://arxiv.org/abs/0706.3531 [nucl-th].

\bibitem{Dumitru:2008wn}
A.~Dumitru, F.~Gelis, L.~McLerran and R.~Venugopalan,
Nucl. Phys. A \textbf{810} (2008), 91-108
https://arxiv.org/abs/0804.3858 [hep-ph].

\bibitem{Gavin:2008ev}
S.~Gavin, L.~McLerran and G.~Moschelli,
Phys. Rev. C \textbf{79} (2009), 051902
https://arxiv.org/abs/0806.4718 [nucl-th].

\bibitem{Takahashi:2009na}
J.~Takahashi, B.~M.~Tavares, W.~L.~Qian, R.~Andrade, F.~Grassi, Y.~Hama, T.~Kodama and N.~Xu,
Phys. Rev. Lett. \textbf{103} (2009), 242301
https://arxiv.org/abs/0902.4870 [nucl-th].

\bibitem{Nagle:2009wr}
J.~L.~Nagle,
Nucl. Phys. A \textbf{830} (2009), 147C-154C
https://arxiv.org/abs/0907.2707 [nucl-ex].

\bibitem{Alver:2010dn}
B.~H.~Alver, C.~Gombeaud, M.~Luzum and J.~Y.~Ollitrault,
Phys. Rev. C \textbf{82} (2010), 034913
https://arxiv.org/abs/1007.5469 [nucl-th].

\bibitem{Petersen:2010cw}
H.~Petersen, G.~Y.~Qin, S.~A.~Bass and B.~Muller,
Phys. Rev. C \textbf{82} (2010), 041901
https://arxiv.org/abs/1008.0625 [nucl-th].

\bibitem{Schenke:2010rr}
B.~Schenke, S.~Jeon and C.~Gale,
Phys. Rev. Lett. \textbf{106} (2011), 042301
https://arxiv.org/abs/1009.3244 [hep-ph].

\bibitem{CMS:2010ifv}
V.~Khachatryan \textit{et al.} [CMS],
JHEP \textbf{09} (2010), 091
https://arxiv.org/abs/1009.4122 [hep-ex].

\bibitem{Skands:2014pea}
P.~Skands, S.~Carrazza and J.~Rojo,
Eur. Phys. J. C \textbf{74} (2014) no.8, 3024
https://arxiv.org/abs/1404.5630 [hep-ph].

\bibitem{Giacalone:2020ymy}
G.~Giacalone,
``A matter of shape: seeing the deformation of atomic nuclei at high-energy colliders,''
PhD thesis, Paris-Saclay University (2020)
https://arxiv.org/abs/2101.00168 [nucl-th].

\bibitem{Qiu:2011iv}
Z.~Qiu and U.~W.~Heinz,
Phys. Rev. C \textbf{84} (2011), 024911
https://arxiv.org/abs/1104.0650 [nucl-th].

\bibitem{Niemi:2012aj}
H.~Niemi, G.~S.~Denicol, H.~Holopainen and P.~Huovinen,
Phys. Rev. C \textbf{87} (2013) no.5, 054901
https://arxiv.org/abs/1212.1008 [nucl-th].

\bibitem{Gardim:2014tya}
F.~G.~Gardim, J.~Noronha-Hostler, M.~Luzum and F.~Grassi,
Phys. Rev. C \textbf{91} (2015) no.3, 034902
https://arxiv.org/abs/1411.2574 [nucl-th].

\bibitem{Teaney:2012ke}
D.~Teaney and L.~Yan,
Phys. Rev. C \textbf{86} (2012), 044908
https://arxiv.org/abs/1206.1905 [nucl-th].

\bibitem{Gardim:2011xv}
F.~G.~Gardim, F.~Grassi, M.~Luzum and J.~Y.~Ollitrault,
Phys. Rev. C \textbf{85} (2012), 024908
https://arxiv.org/abs/1111.6538 [nucl-th].

\bibitem{Gardim:2020mmy}
F.~G.~Gardim and J.~Y.~Ollitrault,
Phys. Rev. C \textbf{103} (2021) no.4, 044907
https://arxiv.org/abs/2010.11919 [nucl-th].

\bibitem{Yan:2015jma}
L.~Yan and J.~Y.~Ollitrault,
Phys. Lett. B \textbf{744} (2015), 82-87
https://arxiv.org/abs/1502.02502 [nucl-th].

\bibitem{ALICE:2017fcd}
S.~Acharya \textit{et al.} [ALICE],
Phys. Lett. B \textbf{773} (2017), 68-80
https://arxiv.org/abs/1705.04377 [nucl-ex].

\bibitem{ALICE:2020sup}
S.~Acharya \textit{et al.} [ALICE],
JHEP \textbf{05} (2020), 085
https://arxiv.org/abs/2002.00633 [nucl-ex].

\bibitem{STAR:2020gcl}
J.~Adam \textit{et al.} [STAR],
Phys. Lett. B \textbf{809} (2020), 135728
https://arxiv.org/abs/2006.13537 [nucl-ex].

\bibitem{STAR:2015mki}
L.~Adamczyk \textit{et al.} [STAR],
Phys. Rev. Lett. \textbf{115} (2015) no.22, 222301
https://arxiv.org/abs/1505.07812 [nucl-ex].

\bibitem{Giacalone:2021udy}
G.~Giacalone, J.~Jia and C.~Zhang,
Phys. Rev. Lett. \textbf{127} (2021) no.24, 242301
https://arxiv.org/abs/2105.01638 [nucl-th].

\bibitem{Schenke:2014tga}
B.~Schenke, P.~Tribedy and R.~Venugopalan,
Phys. Rev. C \textbf{89} (2014) no.6, 064908
https://arxiv.org/abs/1403.2232 [nucl-th].

\bibitem{ALICE:2018lao}
S.~Acharya \textit{et al.} [ALICE],
Phys. Lett. B \textbf{784} (2018), 82-95
https://arxiv.org/abs/1805.01832 [nucl-ex].

\bibitem{CMS:2019cyz}
A.~M.~Sirunyan \textit{et al.} [CMS],
Phys. Rev. C \textbf{100} (2019) no.4, 044902
https://arxiv.org/abs/1901.07997 [hep-ex].

\bibitem{ATLAS:2019dct}
G.~Aad \textit{et al.} [ATLAS],
Phys. Rev. C \textbf{101} (2020) no.2, 024906
https://arxiv.org/abs/1911.04812 [nucl-ex].

\bibitem{STAR:2021mii}
M.~Abdallah \textit{et al.} [STAR],
Phys. Rev. C \textbf{105} (2022) no.1, 014901
https://arxiv.org/abs/2109.00131 [nucl-ex].

\bibitem{Zhang:2021kxj}
C.~Zhang and J.~Jia,
Phys. Rev. Lett. \textbf{128} (2022) no.2, 022301
https://arxiv.org/abs/2109.01631 [nucl-th].

\bibitem{Rong:2022qez}
Y.~T.~Rong, X.~Y.~Wu, B.~N.~Lu and J.~M.~Yao,
Phys. Lett. B \textbf{840} (2023), 137896
https://arxiv.org/abs/2201.02114 [nucl-th].

\bibitem{Ryssens:2023fkv}
W.~Ryssens, G.~Giacalone, B.~Schenke and C.~Shen,
Phys. Rev. Lett. \textbf{130} (2023) no.21, 212302
https://arxiv.org/abs/2302.13617 [nucl-th].

\bibitem{Xu:2021vpn}
H.~j.~Xu, H.~Li, X.~Wang, C.~Shen and F.~Wang,
Phys. Lett. B \textbf{819} (2021), 136453
https://arxiv.org/abs/2103.05595 [nucl-th].

\bibitem{Jia:2022qgl}
J.~Jia, G.~Giacalone and C.~Zhang,
Phys. Rev. Lett. \textbf{131} (2023) no.2, 022301
https://arxiv.org/abs/2206.10449 [nucl-th].

\bibitem{Giacalone:2023cet}
G.~Giacalone, G.~Nijs and W.~van der Schee,
https://arxiv.org/abs/2305.00015   [nucl-th].

\bibitem{PREX:2021umo}
D.~Adhikari \textit{et al.} [PREX],
Phys. Rev. Lett. \textbf{126} (2021) no.17, 172502
https://arxiv.org/abs/2102.10767  [nucl-ex].

\bibitem{Bernhard:2016tnd}
J.~E.~Bernhard, J.~S.~Moreland, S.~A.~Bass, J.~Liu and U.~Heinz,
Phys. Rev. C \textbf{94} (2016) no.2, 024907
https://arxiv.org/abs/1605.03954 [nucl-th].

\bibitem{Nijs:2020roc}
G.~Nijs, W.~van der Schee, U.~G\"ursoy and R.~Snellings,
Phys. Rev. C \textbf{103} (2021) no.5, 054909
https://arxiv.org/abs/2010.15134 [nucl-th].

\bibitem{Bally:2022vgo}
B.~Bally, J.~D.~Brandenburg, G.~Giacalone, U.~Heinz, S.~Huang, J.~Jia, D.~Lee, Y.~J.~Lee, W.~Li and C.~Loizides, \textit{et al.}
https://arxiv.org/abs/2209.11042 [nucl-ex].

\bibitem{STAR:2002hbo}
C.~Adler \textit{et al.} [STAR],
Phys. Rev. C \textbf{66} (2002), 034904
https://arxiv.org/abs/nucl-ex/0206001 [nucl-ex].

\bibitem{Bhalerao:2014xra}
R.~S.~Bhalerao, J.~Y.~Ollitrault and S.~Pal,
Phys. Lett. B \textbf{742} (2015), 94-98
https://arxiv.org/abs/1411.5160 [nucl-th].

\bibitem{Mehrabpour:2018kjs}
H.~Mehrabpour and S.~F.~Taghavi,
Eur. Phys. J. C \textbf{79} (2019) no.1, 88
https://arxiv.org/abs/1805.04695 [nucl-th].

\bibitem{Borghini:2000sa}
N.~Borghini, P.~M.~Dinh and J.~Y.~Ollitrault,
Phys. Rev. C \textbf{63} (2001), 054906
https://arxiv.org/abs/nucl-th/0007063 [nucl-th].

\bibitem{PHENIX:2003qra}
S.~S.~Adler \textit{et al.} [PHENIX],
Phys. Rev. Lett. \textbf{91} (2003), 182301
https://arxiv.org/abs/nucl-ex/0305013 [nucl-ex].

\bibitem{ATLAS:2019peb}
M.~Aaboud \textit{et al.} [ATLAS],
JHEP \textbf{01} (2020), 051
https://arxiv.org/abs/1904.04808 [nucl-ex].

\bibitem{Bhalerao:2006tp}
R.~S.~Bhalerao and J.~Y.~Ollitrault,
Phys. Lett. B \textbf{641} (2006), 260-264
https://arxiv.org/abs/nucl-th/0607009 [nucl-th].

\bibitem{Voloshin:2007pc}
S.~A.~Voloshin, A.~M.~Poskanzer, A.~Tang and G.~Wang,
Phys. Lett. B \textbf{659} (2008), 537-541
https://arxiv.org/abs/0708.0800 [nucl-th].

\bibitem{ALICE:2011ab}
K.~Aamodt \textit{et al.} [ALICE],
Phys. Rev. Lett. \textbf{107} (2011), 032301
https://arxiv.org/abs/1105.3865 [nucl-ex].

\bibitem{ATLAS:2014qxy}
G.~Aad \textit{et al.} [ATLAS],
Eur. Phys. J. C \textbf{74} (2014) no.11, 3157
https://arxiv.org/abs/1408.4342 [hep-ex].

\bibitem{CMS:2022umz}
 CMS Collaboration,
``Probing hydrodynamics and the moments of the elliptic flow distribution in $\sqrt{s_{\mathrm{NN}}}=5.02~\mathrm{TeV}$ lead-lead collisions using higher-order cumulants,'' (2022)
https://cds.cern.ch/record/2806158/files/HIN-21-010-pas.pdf

\bibitem{Giacalone:2016eyu}
G.~Giacalone, L.~Yan, J.~Noronha-Hostler and J.~Y.~Ollitrault,
Phys. Rev. C \textbf{95} (2017) no.1, 014913
https://arxiv.org/abs/1608.01823 [nucl-th].

\bibitem{CMS:2017glf}
A.~M.~Sirunyan \textit{et al.} [CMS],
Phys. Lett. B \textbf{789} (2019), 643-665
https://arxiv.org/abs/1711.05594 [nucl-ex].

\bibitem{ALICE:2018rtz}
S.~Acharya \textit{et al.} [ALICE],
JHEP \textbf{07} (2018), 103
https://arxiv.org/abs/1804.02944 [nucl-ex].

\bibitem{Bhalerao:2018anl}
R.~S.~Bhalerao, G.~Giacalone and J.~Y.~Ollitrault,
Phys. Rev. C \textbf{99} (2019) no.1, 014907
https://arxiv.org/abs/1811.00837 [nucl-th].

\bibitem{NA49:2003njx}
C.~Alt \textit{et al.} [NA49],
Phys. Rev. C \textbf{68} (2003), 034903
https://arxiv.org/abs/nucl-ex/0303001 [nucl-ex].

\bibitem{ATLAS:2014ndd}
G.~Aad \textit{et al.} [ATLAS],
Phys. Rev. C \textbf{90} (2014) no.2, 024905
https://arxiv.org/abs/1403.0489 [hep-ex].

\bibitem{ALICE:2016kpq}
J.~Adam \textit{et al.} [ALICE],
Phys. Rev. Lett. \textbf{117} (2016), 182301
https://arxiv.org/abs/1604.07663 [nucl-ex].

\bibitem{Bozek:2016yoj}
P.~Bozek,
Phys. Rev. C \textbf{93} (2016) no.4, 044908
https://arxiv.org/abs/1601.04513 [nucl-th].

\bibitem{ATLAS:2019pvn}
G.~Aad \textit{et al.} [ATLAS],
Eur. Phys. J. C \textbf{79} (2019) no.12, 985
https://arxiv.org/abs/1907.05176 [nucl-ex].

\bibitem{ALICE:2021gxt}
S.~Acharya \textit{et al.} [ALICE],
Phys. Lett. B \textbf{834} (2022), 137393
https://arxiv.org/abs/2111.06106 [nucl-ex].

\bibitem{ATLAS:2022dov}
G.~Aad \textit{et al.} [ATLAS],
Phys. Rev. C \textbf{107} (2023) no.5, 054910
https://arxiv.org/abs/2205.00039 [nucl-ex].

\bibitem{Giacalone:2018apa}
G.~Giacalone,
Phys. Rev. C \textbf{99} (2019) no.2, 024910
https://arxiv.org/abs/1811.03959 [nucl-th].

\bibitem{Jia:2021tzt}
J.~Jia,
Phys. Rev. C \textbf{105} (2022) no.1, 014905
https://arxiv.org/abs/2106.08768 [nucl-th].

\bibitem{Jia:2021qyu}
J.~Jia,
Phys. Rev. C \textbf{105} (2022) no.4, 044905
https://arxiv.org/abs/2109.00604 [nucl-th].

\bibitem{Bally:2021qys}
B.~Bally, M.~Bender, G.~Giacalone and V.~Som\`a,
Phys. Rev. Lett. \textbf{128} (2022) no.8, 082301
https://arxiv.org/abs/2108.09578 [nucl-th].

\bibitem{Bally:2023dxi}
B.~Bally, G.~Giacalone and M.~Bender,
Eur. Phys. J. A \textbf{59} (2023) no.3, 58
https://arxiv.org/abs/2301.02420 [nucl-th].
\end{thebibliography}

\section*{Acknowledgments} 
This article is based on two talks I gave, first at the Symposium on collective flow in nuclear matter (a celebration of Art Poskanzer's life and career) in December 2022 at Berkeley, then in January 2023 under the program ``Intersection of nuclear structure and high-energy nuclear collisions'' hosted by the Institute for Nuclear Theory at the University of Washington, which I thank for support. 
I thank Giuliano Giacalone for useful input and detailed comments on the manuscript, and Johanna Stachel and Peter Braun-Munzinger for comments on the first version. 
This work is supported by the GLUODYNAMICS project funded by the ``P2IO LabEx (ANR-10-LABX-0038)'' in the framework ``Investissements d'Avenir'' (ANR-11-IDEX-0003-01) managed by the Agence Nationale de la Recherche (ANR), France.

\end{document}